\documentclass[preprint,superscriptaddress,preprintnumbers,amsmath,amssymb
]{revtex4}

\usepackage{graphicx}% Include figure files
\usepackage[dvipsnames]{xcolor}
\usepackage{gensymb}
\usepackage{dcolumn}% Align table columns on decimal point
\usepackage{bm}% bold math
\usepackage{float}
\usepackage{multirow}
\usepackage{amsmath,amssymb}
\usepackage{siunitx}
\usepackage{tabularx}
\usepackage{makecell}
\usepackage{soul}
\usepackage{subcaption}
\usepackage[normalem]{ulem} 

\begin{document}

\title{\textbf{{Photoelectron Spectroscopy Study of U-Te Thin Films: A Unified Perspective of Hybridization Effects across Compositions}}
}% 

\author{E. A. Tereshina-Chitrova}
    \affiliation{Institute of Physics, Czech Academy of Sciences, 18121 Prague, Czech Republic}

\author{S. G. Alex}
    \affiliation{Institute of Physics, Czech Academy of Sciences, 18121 Prague, Czech Republic}

\author{O. Koloskova}
    \affiliation{Institute of Physics, Czech Academy of Sciences, 18121 Prague, Czech Republic}
 
\author{L. Havela}
    \affiliation{Faculty of Mathematics and Physics, Charles University, Czech Republic}

\author{L. Horak}
    \affiliation{Faculty of Mathematics and Physics, Charles University, Czech Republic}
    
\author{O. Romanyuk}
    \affiliation{Institute of Physics, Czech Academy of Sciences, 18121 Prague, Czech Republic}

\author{F. Huber}
    \affiliation{European Commission, Joint Research Centre (JRC), Postfach 2340, DE-76125 Karlsruhe, Germany}

\author{T. Gouder}
    \affiliation{European Commission, Joint Research Centre (JRC), Postfach 2340, DE-76125 Karlsruhe, Germany}

\author{M. Divi\v{s}}
    \affiliation{Faculty of Mathematics and Physics, Charles University, Czech Republic}

\date{\today}

\begin{abstract}
Uranium tellurides span magnetic and superconducting ground states, yet systematic electronic-structure information across the U–Te series remains scarce. In this study, we perform photoemission measurements on freshly prepared U$_x$Te$_y$ thin films covering the range of bulk stoichiometries under ultra-high vacuum (10$^{-9}$ Pa), enabling clean surfaces and compositions matching bulk phases, including the celebrated UTe$_2$. X-ray and ultraviolet photoelectron spectroscopy (XPS/UPS) reveal consistent evolution of the U 4\textit{f} and Te 3\textit{d} core levels and valence states across the series, in good agreement with the limited bulk data. Supported by uniform \textit{ab initio} calculations for all U–Te compounds, we identify systematic trends in U–Te hybridization and charge-transfer effects across the series. These results establish thin-film photoemission as a reliable route for mapping electronic-structure trends in tellurides of heavy elements with diverse electronic ground states.
\end{abstract}


\maketitle

%\tableofcontents

\section{\label{sec:level1}Introduction}

The physics of actinides is remarkably complex due to intricate interactions involving their spatially extended 5\textit{f} electron shells \cite{sech}. This complexity is driven by relativistic effects, strong electron correlations, and significant hybridization both within the actinide electronic states and between these states and ligand orbitals. These features, particularly pronounced in early actinides, lead to intricate bonding environments, posing substantial challenges for reliable modeling of their electronic structures. When combined with chalcogen ligands such as tellurium, this complexity may be further enriched, as tellurides are known to adopt a wide range of bonding motifs~\cite{Steinberg2025}. These challenges are particularly evident in the unconventional superconductor UTe$_2$, a material recently discovered to exhibit exceptional tolerance to high magnetic fields and potential for topological superconductivity \cite{ran,Aoki_2022}. Although there is a general consensus that UTe$_2$ exhibits intermediate 5\textit{f} occupancy between 5\textit{f}$^{2}$ (U$^{4+}$) and 5\textit{f}$^{3}$ (U$^{3+}$) states, its exact electron configuration has been the subject of considerable debate fueled by diverse experimental \cite{fujimori2019,wilhelm2023,Miao,christovam,liu} and theoretical \cite{shick2021,Miao,sergei,shick2024} findings. Besides the fact that different spectroscopic techniques have varying sensitivities to different aspects of 5\textit{f} electron behavior, the interpretation of experimental results often relies \cite{fujimori2016} on comparative studies of uranium-based compounds with diverse ligand types and crystal structures, further complicating conclusions about 5\textit{f} electron behavior.

Indeed, factors such as atomic arrangement, coordination, electron configurations, and ligand electronegativities influence charge redistribution, yielding a wide range of physical properties among these materials. The variability is well illustrated in the U\textit{X}$_2$ series with \textit{p}-block elements (\textit{X} = Ge, Te, Ga). For instance, UGa$_2$ crystallizes in a hexagonal structure ($P6/mmm$) and has a Curie temperature of 125 K \cite{andreev1979}, exhibiting ordered magnetic moments near 3 $\mu$$_B$ per U atom. In contrast, the orthorhombic UTe$_2$ ($Immm$) remains a paramagnetic superconductor with an effective magnetic moment of 2.8 $\mu$$_B$/U \cite{ran}. UGe$_2$ ($Cmmm$) is a ferromagnet (and a superconductor under pressure) with a lower ordered moment of 1.42 $\mu$$_B$/U and a Curie temperature of 52 K \cite{BOULET1997104}. Given these discrepancies, a thorough investigation of hybridization effects within a consistent ligand family is highly desirable. 

Among these ligand classes, tellurium occupies a particularly flexible role: its ability to support diverse binding environments~\cite{Steinberg2025} underlies the strong dependence of the U–Te electronic structure on stoichiometry and local atomic arrangement, providing a suitable platform for probing 5\textit{f}–ligand hybridization. Here, we systematically explore a series of uranium tellurides using a thin-film approach that enables continuous tuning of composition, targeting stoichiometries corresponding to bulk phases in the U–Te phase diagram \cite{okamoto1993}. Across the series, distinct stoichiometries and crystal structures give rise to markedly different electronic ground states, ranging from ferromagnetism to superconductivity. Moreover, limited thermodynamic stability of some of the U–Te compounds (for instance, U$_3$Te$_4$) has hindered their experimental characterization in bulk form \cite{landolt}, an issue that can be partially alleviated by thin-film growth under controlled conditions. Prior to the discovery of exotic superconductivity in UTe$_2$, UTe was the only uranium telluride studied in greater detail. It is a local-moment ferromagnet that exhibits a pronounced increase of the Curie temperature under hydrostatic pressure from 104 K to 181 K at 7.5 GPa \cite{Link_1992}, an effect attributed to changes in the U–Te hybridization \cite{COOPER1994120}.

The present study represents a pioneering effort to establish a consistent spectroscopy map across U–Te stoichiometries under identical preparation conditions, enabled by a thin-film approach. The article is organized as follows: Section II outlines our experimental and computational methodologies. Section III presents in situ X-ray and ultraviolet photoelectron spectroscopy (XPS and UPS) data acquired under UHV conditions on pristine, freshly prepared thin-film surfaces, enabling a systematic comparison of composition-dependent shifts and line-shape changes across the U–Te series. Supporting X-ray diffraction (XRD) data are provided to document the structural/phase assignment of the films. We also conducted magnetic and resistivity studies across all compositions; however, these findings will be elaborated in subsequent publications to maintain the focus of the present communication.

In Section IV, we discuss bonding trends by estimating potential charge transfer across the investigated U–Te stoichiometries, supported by first-principles Density Functional Theory (DFT) calculations performed consistently for all phases. The theoretical results are discussed in direct connection with the experimental electronic-structure trends revealed by photoelectron spectroscopy. This comparative multi-composition approach provides a consistent basis for understanding hybridization phenomena in uranium tellurides.

\section{\label{sec:level1}Materials and methods}

\subsection{\label{sec:level2}Samples synthesis} Thin film samples of various U-Te compositions containing less than 1 at.\% of carbon and oxygen (see survey spectra in Appendix~A, Fig. A1) were synthesized using a home-built ultra-high vacuum (UHV) (base pressure 10$^{-9}$ Pa) triode DC sputtering equipment suitable for the synthesis of various alloy films \cite{U-Mo,PAUKOV2018113,tereshina2023}. For the fabrication of U-Te thin films, separate U (natural uranium, 99.9 wt.\% purity) and Te (purity 5N) targets were used. Purified Ar served as the sputter gas, with the Ar pressure held constant at 5.5 × 10$^{-1}$ Pa during deposition. The sputtering discharge current on the U target was maintained at 2 mA. Plasma stability was maintained by an electron-emitting thoriated tungsten filament. Because Te exhibits a high vapor pressure (0.1 Pa at 277 \textdegree C), its controlled deposition is challenging. Te was therefore evaporated from a confined crucible geometry with a small aperture, allowing the Te flux to mix with the U sputtered flux before reaching the substrate. The intended stoichiometry of the films was achieved by controlling adjustable parameters such as the temperature of the Te target and the negative DC voltage (-790 V) applied to the uranium target. 

We primarily used Si wafers with (001) orientation as substrates. Single-crystalline MgO, LiNbO$_3$, Al$_2$O$_3$, and CaF$_2$ substrates were also employed in some growth trials, particularly for UTe$_2$, where a possibility of lattice matching was explored. The Si substrates were Ar-ion sputter-cleaned at $T = 523\,\mathrm{K}$, while other substrates were cleaned by annealing at high temperatures ($500{-}700^\circ\mathrm{C}$). 
Deposition was performed at room temperature to minimize film--substrate interactions. The films were intentionally kept 10--30~nm thick, which corresponds to the probing depth of the surface-sensitive techniques employed later and ensures continuous films suitable for structural and compositional studies. Under these conditions, the arriving U and Te fluxes determine the film stoichiometry, which is therefore substrate-independent. To relate composition to structure, matched sets of films were grown under identical deposition conditions: the Si-based films were kept under UHV for subsequent analyses, while parallel films grown in the same runs on crystalline substrates were taken out for \emph{ex-situ} XRD. Representative diffraction patterns of the crystalline-substrate films are shown in Appendix~A, Fig. A2.

\subsection{\label{sec:level2}Spectroscopy methods} For spectroscopic analysis, the films were transferred (in 2-3 min) from the deposition chamber to the surface analysis chamber under UHV conditions (without breaking vacuum), ensuring pristine surfaces for examination. The spectrometer uses monochromated Al–K$\alpha$ radiation (h$\nu$ = 1486.6 eV) for XPS and He II and He I UV radiation (40.81 eV and 21.22 eV, respectively) for UPS. SPECS PHOIBOS 150 MCD-9 was used as the  electron energy analyzer. Calibration of the spectrometer was performed using the Au 4\textit{f}$_{7/2}$ line (83.9 eV binding energy) and the Cu 2\textit{p}$_{3/2}$ line (932.7 eV binding energy) for XPS, along with the Au Fermi level for UPS. The samples were electrically grounded during measurements, and no sample charging was observed in either XPS or UPS. Overview XPS spectra, collected to prove the absence of spurious elements, were supplemented by probing the O 1\textit{s} and C 1\textit{s} energy windows in detail. The stoichiometry was determined by taking the ratio of the integrated intensities of the U 4\textit{f} and Te 3\textit{d} core-level lines, calculated after the Shirley background subtraction. A consistent energy window was used throughout all compositions to ensure comparability. The integrated intensities were then adjusted using empirical sensitivity factors provided by SPECS \cite{SPECSRev9}. The precision of XPS-derived U:Te ratios is approximately $5$\% and the absolute accuracy checked by subsequent Rutherford Backscattering
Spectrometry (RBS, will be shown elsewhere) is better than 10\%. Since all spectra were acquired under identical geometry and comparable film thickness, the XPS-derived composition is used primarily as a consistent relative measure across the series. The energy resolution achieved in the UPS mode is typically around 60 meV. 

In addition, for tellurium we also measured bulk valence-band reference spectrum (see comparison of the UPS and XPS data in Appendix~A, Fig. A3) using a Kratos Axis Supra photoelectron spectrometer (Al~K$\alpha$, X-ray power of 150~W, pass energy of 40~eV). The surface of the Te sample was cleaned by gas-cluster ion-beam sputtering (GCIS) using an Ar$^{+}$ cluster beam with an energy of 5~keV, cluster size of 1000 ions, sputter crater area of $2 \times 2~\mathrm{mm}^{2}$, and a sputtering time of 30~min. No carbon contamination layer was detected on the Te surface. Figure A3 also compares these reference data with the UPS spectrum (h$\nu$ = 40.81 eV) of the thin-film sample, obtained using the same PHOIBOS analyzer and acquisition conditions as used for the other films in this study.

\subsection{\label{sec:level2}Computational methods} To estimate partial charges at U and Te atomic positions, we employed the Full Potential Local Orbitals (FPLO) \cite{Koepernik} and general potential (Linearized) Augmented Plane Wave plus local orbitals ((L)APW + lo) methods \cite{SCHWARZ200271} to calculate electronic structure and material properties for various U-Te bulk compositions shown in Table I from first principles based on Density Functional Theory (DFT). Although more advanced techniques (as DFT+\textit{U}) would be necessary to account for electron-electron correlations and reproduce correctly possible band gap, our computational strategy was to avoid any external parameters and to use scalar relativistic DFT in LSDA approximation to assess bonding variations between various U-Te species, the structures of which were taken at experimental equilibrium reported in literature. The total electronic density of states is fully comparable for both FPLO and (L)APW + lo methods. The partial electronic charges were calculated by FPLO method to obtain the systematic uranium bonding properties of the bulk U-Te compositions. The results of our calculations are provided in Table II.

\section{\label{sec:level1}Experimental results}

\subsection{\label{sec:level2}X-ray Photoelectron Spectroscopy Study}
The U-Te films were previously found to be highly air-sensitive \cite{tereshina2023}, so the primary concern was degradation. We ensured that all spectroscopic studies were conducted on impurity-free surfaces (see Appendix~A, Fig. A1). We investigated the evolution of the U-4\textit{f} (Fig. 1, left) and Te-3\textit{d} core-level spectra (Fig. 1, right) for U-Te thin films with varying Te:U ratios. These spectra were recorded both to determine the stoichiometry of the films and to probe the local electronic structure across U-Te compositions \cite{thompson}. 

For rare earth-based 4\textit{f}-electron systems \cite{UWAMINO198467} and 5\textit{f}-electron actinides such as uranium oxides, where oxidation states are well-characterized due to their ionic nature, core-level XPS is a straightforward tool revealing details of electronic states \cite{Gouder2018}. However, uranium-based materials containing metals or metalloids exhibit significant many-body effects \cite{thompson,fujimori2016}, which complicate the use of conventional spectral quantification techniques \cite{Gouder2002}. In the present work, we conduct a comparative analysis of various U$_x$Te$_y$ compositions in an attempt to uncover systematic trends driven by changes in the local environment.

In metallic uranium, the U-4\textit{f} core-level lines, spin-orbit split by 11 eV, are located at 377.3±0.1 eV and 388.2±0.1 eV binding energy (BE) for the 4\textit{f}$_{7/2}$ and 4\textit{f}$_{5/2}$ components, respectively. Both peaks show pronounced asymmetry with a high-binding-energy tail (Fig.1, left), which arises from electron-electron interaction effects and is characteristic of a high density of electronic states at the Fermi level ($E_{\mathrm{F}}$) \cite{Schneider1981}. Such spectral asymmetry is commonly described by a Doniach–Sunjić-type lineshape and reflects strong metallic screening associated with itinerant 5$f$ electrons, enhancing the density of states at the Fermi level. A similar spectral shape is observed in UTe$_{0.20}$, suggesting that the 5$f$ states in UTe$_{0.20}$ remain delocalized and strongly hybridized with conduction electrons, indicating metallic behavior. 

\begin{figure}[h!]
\centering
\includegraphics[width=0.45\textwidth]{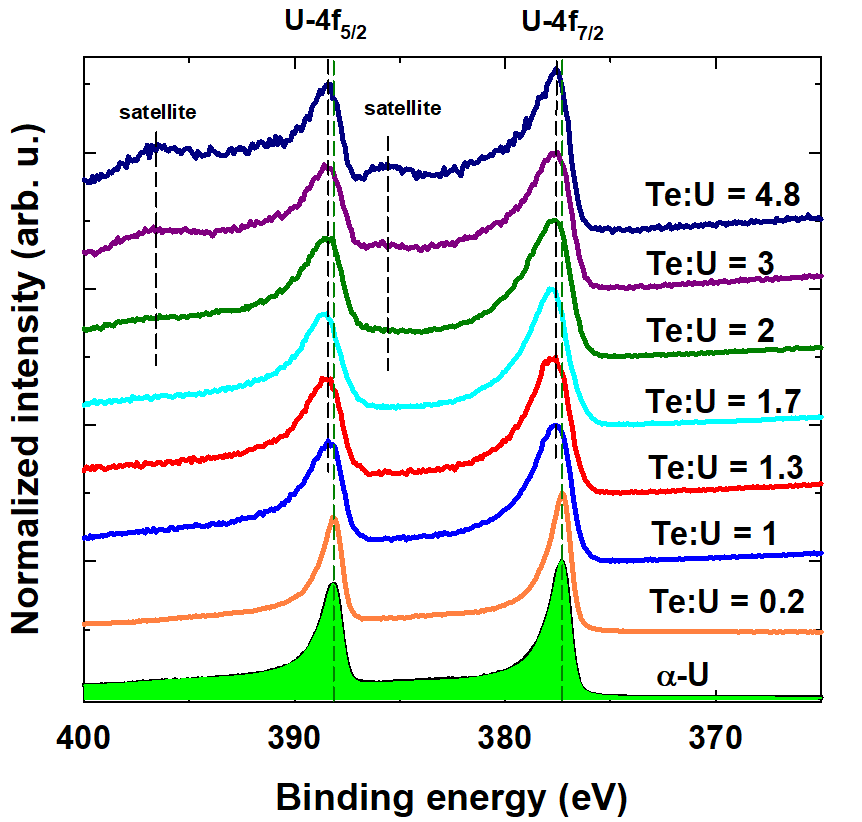}\,
\includegraphics[width=0.45\textwidth]{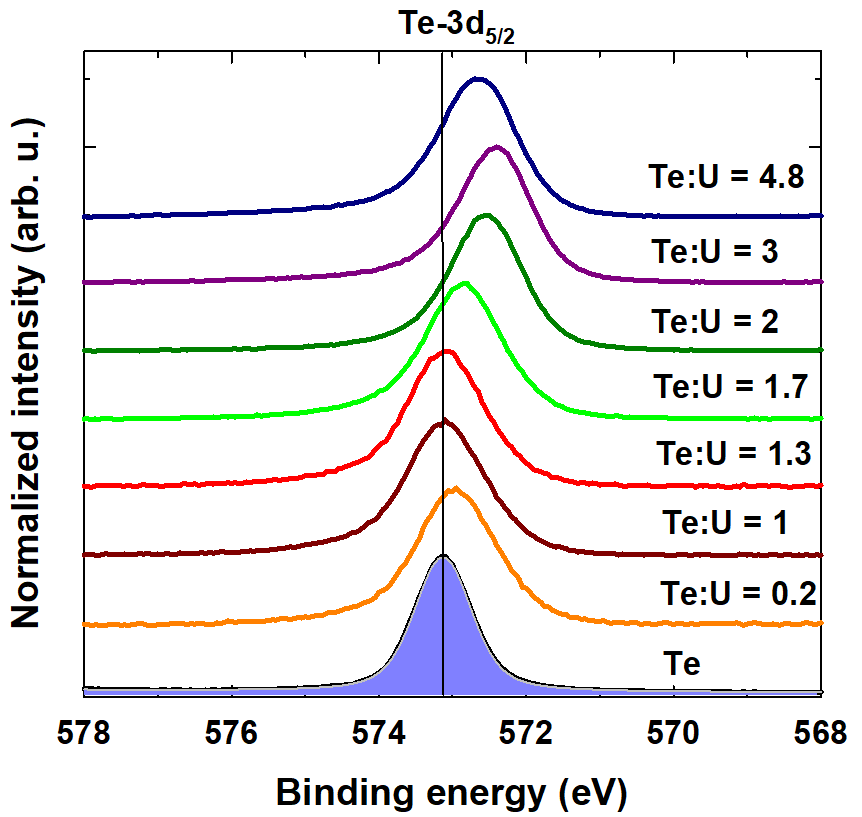}
\caption{(Left) U-4\textit{f} core-level spectra of U-Te thin films, measured for varying Te:U ratios, display the 4\textit{f}$_{7/2}$ and 4\textit{f}$_{5/2}$ peaks, separated by 11 eV due to spin-orbit interaction. (Right) For Te, the evolution of the dominant core-level line Te-3\textit{d}$_{5/2}$ normalized to the maximum of its intensity is shown for various Te:U concentrations (the Te-3\textit{d}$_{3/2}$ line is located at approx. 583 eV).}
\label{fig:2}
\end{figure}

While the uranium core levels practically do not shift in energy, the Te-3\textit{d}$_{5/2}$ line in UTe$_{0.20}$ shows a slight shift to lower binding energy compared to elemental Te, suggesting charge transfer toward Te. This behavior is consistent with previous findings \cite{PAUKOV2018113} showing that alloying without compound formation has only a minor effect on the U core-level lineshape and position. In contrast, spectral changes in XPS typically emerge upon the formation of well-defined compounds with distinct chemical bonding environments.

\begin{figure}[h!]
\centering
\includegraphics[width=0.7\textwidth]{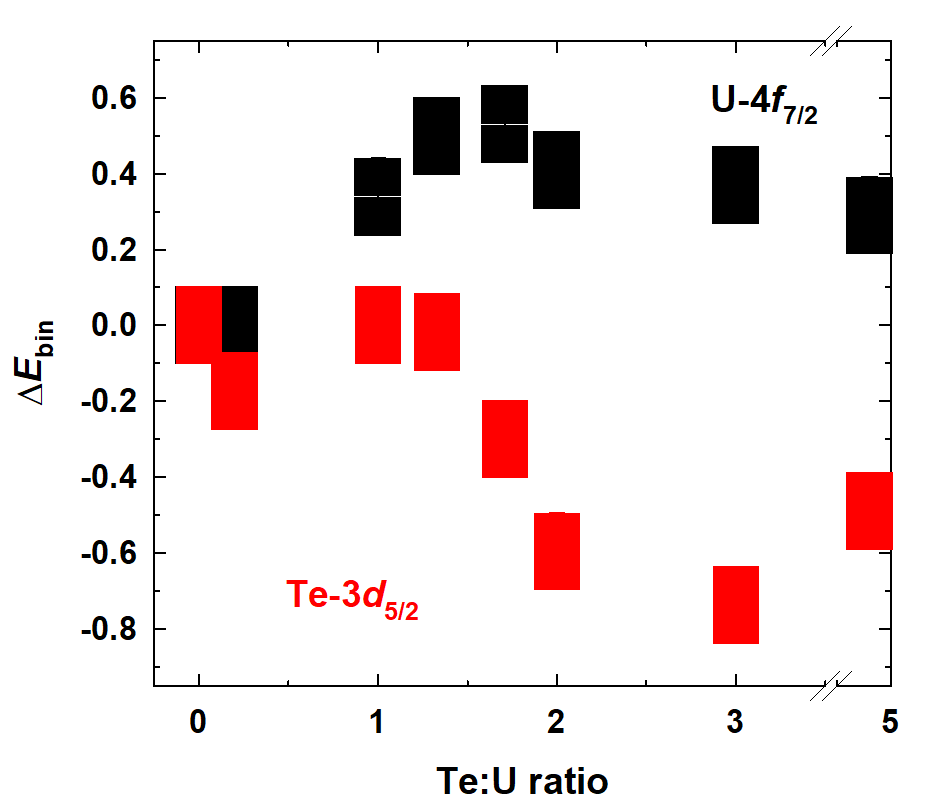}\,
\caption{Binding energy shifts for the dominant U-4\textit{f}$_{7/2}$ and Te-3\textit{d}$_{5/2}$ core-level lines as compared to $\alpha$-U (the reference position is at {377.3}\textpm0.1 eV) and pure Te (the reference position is at {573.1}\textpm0.1 eV) in U-Te films. The size of the data points includes the estimated experimental error.}
\label{fig:3}
\end{figure}

As the Te concentration increases and stoichiometric compounds begin to form, notable changes appear in both the U and Te core-level spectra. In the first compound of the series, UTe, the U-4\textit{f}$_{7/2}$ line broadens significantly (full width at half maximum (FWHM) 2.2 eV vs. 1.7 eV in $\alpha$-U) and both U-4$f$ and Te-3$d$ move to higher binding energy, with the Te-3$d$ maximum returning close to the elemental-Te position. The shift of both U and Te lines can be understood as the effect of chemical bonding reducing the energy of both constituents. An important fact is that the dominant screening of the U-4$f$ core hole by the 5$f$ electrons remains in action, which assumes an itinerant character of the 5$f$ states has to be at least partly retained. The small shift of the maximum points to the onset of charge redistribution from U to Te affecting a majority of U sites. The broadening of the 4$f$ emission lines on the higher binding energy side can be attributed to many body effects. We have to stress that despite extensive prior studies of uranium monochalcogenides and monopnictides \cite{GREUTER1980117,Takeda2009,durak2004,gerry1990,durak2004-1}, no bulk XPS core-level data for UTe are available for direct comparison with the present spectra. Proceeding to higher Te concentrations, the 3:4 phase (Te:U = 1.3 in Fig. 1), familiar from uranium pnictides studies \cite{suga}, shows even broader and more asymmetric U-4$f$ lines attributed to increased 5\textit{f} hybridization and modified core-hole screening.

The Te-3\textit{d}$_{5/2}$ core-level peak displays a non-monotonic evolution with the Te concentration variation (see Fig. 2): the binding energy first increases in compounds with a single Te site (UTe, U$_3$Te$_4$), then gradually shifts to lower values in more Te-rich compounds (\text{Te:U}~$\geq$~1.7),  where multiple inequivalent Te sites are present (Table I). This trend points to a complex interplay between local coordination, U–Te bonding geometries, and core-hole screening, indicative of gradual charge redistribution and evolving U-Te hybridization (see Section IV for details of calculated occupancies of selected electronic states). Although Fujimori et al. observed splitting of the Te-4\textit{d} lines in the UTe$_2$ single crystal (see Appendix~A, Fig. A4), our Te-3\textit{d} spectra exhibit no such resolved features. The moderate broadening we observe already in UTe$_{0.20}$ indicates that structural complexity is not the dominant factor in Te-3\textit{d} peak broadening.

A distinct satellite-like feature emerges on the high-binding-energy side of the U-4\textit{f} peak in UTe$_2$ and more Te-rich compositions. Such satellite structures are characteristic of strongly correlated electron systems and originate from multiple final states in the photoemission process. Fujimori et al. \cite{fujimori2019} attributed the satellite in UTe$_2$ to a mixed-valence ground state. Indeed, variable fractional 5\textit{f} occupancy was suggested on the basis of XAS/XMCD study \cite{wilhelm2023}.

For UTe$_3$, which exists in two structural modifications, we were not able to distinguish between them based on XRD analysis due to microstructure effects; however, magnetization measurements indicate the formation of an antiferromagnetic phase with a Néel temperature of 5 K, consistent with the $\alpha$-UTe$_3$ polymorph (cf. Table I; detailed results will be published elsewhere). The U-4\textit{f} lines do not differ significantly from those observed in UTe$_2$. The Te-3\textit{d}$_{5/2}$ line, however, shows a slight additional shift toward lower binding energies and noticeable broadening, which likely reflects the presence of multiple Te sites.

For the final film in the series, the estimated Te:U stoichiometry of 4.8 is close to the UTe$_5$ phase. At this composition, the uranium signal is already strongly attenuated, bringing a larger uncertainty in the stoichiometry determination. The Te-3\textit{d}$_{5/2}$ line shifts somewhat back to a higher binding energy, consistent with the Te concentration approaching the elemental Te limit. The formation of a Te-rich surface overlayer — similar to that observed in U-Ga films prepared at elevated temperatures \cite{GOUDER20017} can be excluded, as UPS data (cf. Fig. 3 and Appendix~A, Fig. A3) do not support this scenario.

Overall, the evolution of the U-4\textit{f} and Te-3\textit{d} core-level spectra across the U–Te series reflects systematic changes in the electronic structure with increasing tellurium concentration. The emergence of satellite features, shifts in core-level binding energies, and spectral broadening all point to modifications in the U-Te hybridization and charge distribution. Electronic structure calculations in Section IV provide additional insights into the interaction of 5\textit{f} states with ligands.

Besides the U-4\textit{f} and Te-3\textit{d} lines we recorded for some U-Te concentrations also the U-6\textit{p}$_{3/2}$ lines, which are the most shallow U core-level lines (see Fig. 5). The larger spatial extent (compared to the 4\textit{f} states) suggests that the U-6\textit{p} spectra may be much more sensitive to the situation of the 5\textit{f} and other valence-band states of U via the 6\textit{p}-5\textit{f},6\textit{d} Coulomb interaction. A drawback is much lower intensity of the 6\textit{p} spectra, requiring much longer acquisition time. Hence they are recorded rather seldom and there is not sufficient data for a systematic comparison. 

Fig. 5 shows a large difference between $\alpha$-U, exhibiting a double peak at 15.5 and 17.5 eV BE (the splitting is likely due to crystal field effects) and the U-Te compounds. From those, the UTe$_{0.8}$, which likely contains majority of the UTe phase, has the maximum at 17.5 eV, while it shifts to 18.0 eV for higher Te concentrations, reflecting progressive charge redistribution away from U, which reduces the Coulomb repulsion. Such shift to approx. 18 eV BE was found in UH$_3$, resulting from a massive transfer of U-6\textit{d} states towards H \cite{HAVELAU-H3}. It is interesting to point out that the polar character of bonding is also in the U-Te case projected in the 6\textit{p} shifts, whereas the position of the 4\textit{f} lines changes very little, while the variations of the shape are more conspicuous in the latter case. 

\subsection{\label{sec:level2}Ultraviolet Photoelectron Spectroscopy Study}

Details of occupied electronic states in the vicinity of the Fermi level are revealed by UPS using the photon energy of 40.81 eV (He II), which has a substantial photoexcitation cross section for the 5\textit{f} states \cite{Yeh1985}. In addition, comparison with He I spectra (21.22 eV) provides complementary sensitivity to non-\textit{f} valence states \cite{opeil}. The remarkable surface sensitivity of UPS, which probes depths smaller than XPS (with an effective information depth on the order of a few atomic layers at these photon energies), indicates that our samples are free of contaminants, except for the starting U film, which contains up to 5\% oxygen, as indicated by the presence of O-2\textit{p} states at a binding energy of 6 eV. In $\alpha$-U metal, the 5\textit{f} emission exhibits a characteristic triangular profile anchored at the Fermi level ($E_{\mathrm{F}}$), with a peak intensity just below $E_{\mathrm{F}}$ and a tail that extends toward higher binding energies (Fig. 3). The profile is indicative of itinerant 5\textit{f} states, consistent with uranium structural stability under pressures as high as 100 GPa \cite{Akella1997}.

When tellurium is introduced, the uranium is diluted, yet the alloy UTe$_{0.2}$ (which does not correspond to a stable stoichiometric compound) retains a band-like character in its 5\textit{f} states. (Here and below, the stoichiometric ratios for the samples presented were determined from the intensity ratios of U-4\textit{f}$_{7/2}$ and Te-3\textit{d}$_{5/2}$ core-level peaks, as detailed in Section 3.1). Besides the 5\textit{f}-emission close to $E_{\mathrm{F}}$, the broad shoulder observed at 0.5–1 eV in UTe$_{0.2}$ may correspond to a shake-up satellite, suggesting that the system remains in an excited state after the photoelectron has been emitted \cite{koloskova}.

\begin{figure}[h!]
\centering
\includegraphics[width=0.7\textwidth]{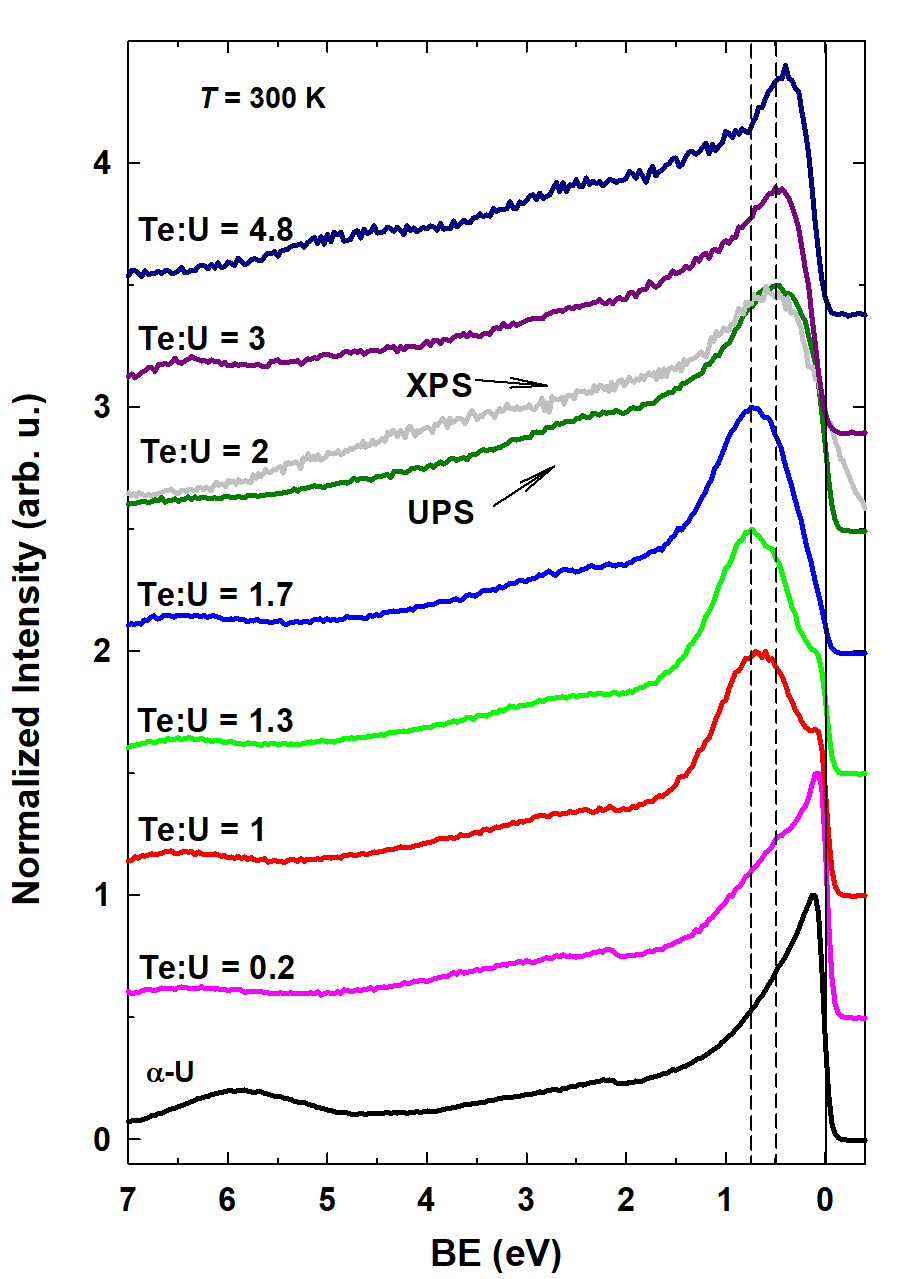}\,
\caption{Normalized He II UPS spectra for the initial $\alpha$-U film compared with various U-Te composition films. For the UTe$_2$ thin film, we also show the XPS data, scaled to match the UPS data. Vertical solid line marks the Fermi energy $E_{\mathrm{F}}$ = 0, while broken vertical lines serve as visual guides.}
\label{fig:3}
\end{figure}

\begin{figure}[h!]
\centering
\includegraphics[width=0.7\textwidth]{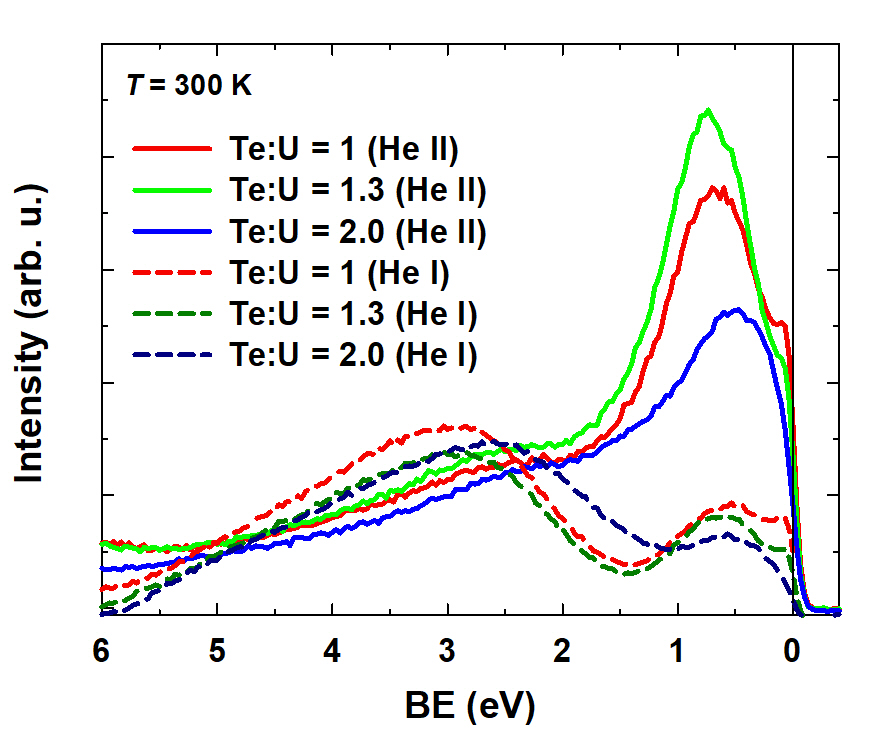}\,
\caption{Comparison of He II with He I UPS spectra for selected U-Te films compositions.}
\label{fig:4}
\end{figure}

\begin{figure}[h!]
\centering
\includegraphics[width=0.7\textwidth]{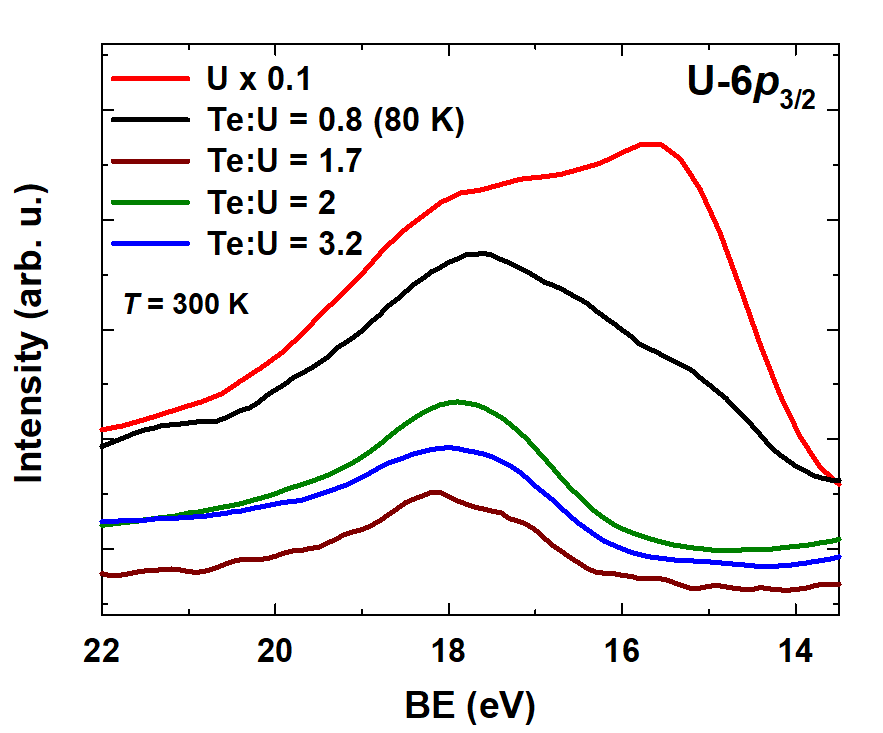}\,
\caption{The evolution of the U-6\textit{p}$_{3/2}$ emission across various U-Te compositions. The U6\textit{p}$_{1/2}$ is located at higher binding energies (25-27 eV).}
\label{fig:4}
\end{figure}

Additionally, the small feature above 2 eV in the spectra of $\alpha$-U is commonly attributed to U 6\textit{d} valence states \cite{reihl1985}. While this feature is sharp in $\alpha$-U and UTe$_{0.2}$, it broadens and shifts to higher binding energies as Te content increases, reflecting overall band widening. This evolution is consistent with a general reduction in U-U distances (Table I) as the Te:U ratio increases, which enhances the hybridization between U-5\textit{f} and 6\textit{d} (and, to a lesser extent, 7\textit{s}) and Te 5\textit{p} orbitals, resulting in a change in bonding.

The intensity of 5\textit{f} emission at $E_{\mathrm{F}}$ decreases with further increasing Te content, marked by the development of a broad maximum at 0.7 eV BE in the UTe film, while a sharp Fermi level cutoff is preserved. This spectrum aligns closely with observations on bulk UTe data \cite{durak2004, reihl1981}. Durakiewicz et al. \cite{durak2004-1} suggested that UTe exhibits both localized and itinerant 5\textit{f} electron behavior, indicative of a complex interplay between these states \cite{FRICK1985549}. The lowered density of states (DOS) at the Fermi level in UTe, relative to $\alpha$-U, aligns with the observed resistivity of 172 \SI{}{\micro\ohm}·cm \cite{FRICK1985549}, despite a comparable electronic specific heat coefficient (Sommerfeld coefficient) of $\gamma$ = 10.3 mJ/mol·K$^2$ (Table I). Our electronic structure calculations (Table II), supported by He I data sensitive to U-6\textit{d} states (Fig. 4), reveal for UTe (in comparison to the more Te-rich compositions) an enhanced U-6\textit{d} charge. This can be understood as a reduced drain of the 6$d$ electrons to ligands due to the lowest Te/U ratio in UTe.

Notably, our thin UTe films also display a negative temperature coefficient of resistivity as the temperature approaches the Curie temperature of approximately 100 K, closely mirroring bulk properties. (Detailed data supporting this observation will be provided in a forthcoming publication.) This correspondence implies that the properties and electronic configurations observed in thin films closely reflect those in bulk materials, corroborating the suitability of thin films as a reliable platform for studying the electronic characteristics of uranium-telluride compounds. 

\textbf{Table I.} Bulk stoichiometric compositions are listed together with available literature data on their physical properties. The table summarizes the crystal structures, Te:U atomic ratios, Sommerfeld coefficient $\gamma$ (given per formula unit), magnetic ground state (PM: paramagnetic; FM: ferromagnetic; AFM: antiferromagnetic), Curie temperature $T_C$ where applicable, and effective magnetic moment $\mu_{\mathrm{eff}}$. The quantities $d_{\text{U--U}}$ and $d_{\text{U--Te}}$ denote the shortest U-U and U-Te interatomic distances, respectively.

\begin{ruledtabular}
\begin{tabular}{cccccccc}
 &Struct. &Te/U &\makecell{$\gamma$\\(mJ/mol·K$^2$)} &Magnetism &\makecell{T$_C$\\(K)} &\makecell{$\mu$$_{eff}$\\($\mu$$_B$)} &\makecell{d$_{U-U}$/d$_{U-Te}$\\(\AA)}\\
\hline
UTe& $Fm\overline{3}m$ & 1 &10.3 \cite{RUDIGIER1983803} & FM & 104 & 2.25 \cite{gerry1990} & 4.36/4.36\\
U$_3$Te$_4$ & $I\overline{4}3d$ & 1.33 &- & FM & 38 \cite{suski1972magnetic} or 105 \cite{chechernikov1967magnetic} & - &4.40/3.51\\
U$_2$Te$_3$ & $Pnma$ & 1.5 &- & FM & 110 & 2.87 \cite{TOUGAIT199867} &4.28/3.60\\
U$_3$Te$_5$ & $Pnma$ & 1.66 &- & FM & 120 & 2.69 \cite{TOUGAIT1998356} &4.12/3.57\\
U$_7$Te$_{12}$& $P\overline{6}$ & 1.71 &48 \cite{Opletal2023}& FM & 48  &1 \cite{Opletal2023} & 3.92/3.58\\
UTe$_2$& $Immm$ & 2 &120-150 \cite{Aoki_2022} & PM & - &2.8 \cite{ran} & 3.78/3.07\\
U$_2$Te$_5$& $C2/m$ & 2.5 &- & PM & - &2.83 \cite{TOUGAIT1997320} & 3.79/2.90\\
$\alpha$-UTe$_3$& $P2_1/m$ & 3 &- & AFM & 5 &3.09 \cite{NOEL1986265} & 4.2/2.64\\
$\beta$-UTe$_3$& $Cmcm$ & 3 &130 \cite{thomas2025enhancedtwodimensionalferromagnetismvan} &  FM & 15 &3.6 \cite{thomas2025enhancedtwodimensionalferromagnetismvan} & 4.34/3.07\\
UTe$_5$& $Pnma$ & 5 &- &  weak FM & 8 &3.34 \cite{NOEL19841171} & 4.22/2.81\\
\end{tabular}
\end{ruledtabular}

For the film with the Te:U ratio 1.3, corresponding to the bulk composition U$_3$Te$_4$, we observe a further decrease of the emission at the Fermi level (see Fig. 3). The bulk U$_3$Te$_4$ has been studied only seldom, with its magnetic structure and Curie temperature remaining subjects of controversy in the literature \cite{suski1972magnetic,chechernikov1967magnetic}. Resistivity measurements suggest that U$_3$Te$_4$ behaves as a semimetallic conductor \cite{BLAISE1981417}. The observed reduction of the  spectral weight at $E_{\mathrm{F}}$ aligns with expectations for a semimetal. Interestingly, our calculations (next section) reveal that the 6\textit{d} orbitals in U$_3$Te$_4$ are significantly impacted by bonding with Te, leaving the 5\textit{f} orbitals less involved in the bonding. This is reflected in the enhanced shoulder structure at \(\sim 0.7 \, \mathrm{eV}\) in U\textsubscript{3}Te\textsubscript{4} compared to UTe (see Fig. 3).  We compared our results with available literature photoemission spectroscopy data of single-crystalline U\textsubscript{3}P\textsubscript{4} and U\textsubscript{3}As\textsubscript{4}, which share the same crystal structure type Th\textsubscript{3}P\textsubscript{4} but have lattice parameters by \(\sim10\% \) smaller than in U\textsubscript{3}Te\textsubscript{4}. These studies \cite{SUGA1985297} using synchrotron radiation with \(h\nu = 32{-}140 \, \mathrm{eV}\) revealed very similar behavior to what we observed here.

With an additional increase in Te concentration to Te:U = 1.7, we observe further DOS reduction at $E_{\mathrm{F}}$ relative to the metallic state of $\alpha$-U, with the maximum at 0.7 eV becoming a predominant spectral feature. Currently there are no photoelectron data in the literature to compare with the corresponding bulk phases U$_3$Te$_5$ and U$_7$Te$_{12}$, which are close to the film composition we obtained (Te:U = 1.7$\pm$0.1). Nevertheless, the resistivity study for U$_7$Te$_{12}$ suggests a "half-gapped" semimetallic state \cite{Opletal2023}, a description consistent with opening the pseudogap in our data. The low \textit{N}($E_{\mathrm{F}}$) observed in our spectra contrasts with the appreciable electronic specific heat coefficient $\gamma$ = 48 mJ/mol·K$^2$ for U$_7$Te$_{12}$ \cite{Opletal2023}. The observed situation suggests, on one hand, the presence of rather localized \textit{f}-electrons that weakly contribute to conduction, but  enhance \(\gamma\) by strengthened electron-electron correlations.

For the Te:U = 2 film, UTe$_2$, the main emission peak is broadened and shifted somewhat towards the Fermi level, its maximum being at around 0.5 eV. The decrease of intensity on approach to $E_{\mathrm{F}}$ is interrupted by a sharper Fermi level cutoff (Figs. 3 and 4), shaped by the Fermi-Dirac distribution function, confirming the metallic character of the film. The peak position agrees with findings from angle-resolved photoemission studies on UTe$_2$ single crystals by Miao et al. \cite{Miao}, for which the DFT+DMFT calculations associate this feature predominantly with 5\textit{f}$^2$ multiplet excitations, while the sharp spectral weight at the Fermi level is attributed to a Kondo-like quasiparticle resonance arising from hybridization between 5$f$ states and conduction electrons. In contrast, Fujimori et al. \cite{fujimori2019}, using soft x-ray synchrotron radiation (h$\nu$ = 800 eV), reported stronger 5$f$ spectral intensity directly at the Fermi level, characteristic of more itinerant 5\textit{f}$^3$-like behavior. However, their data also display a broad incoherent structure at higher binding energies, indicating strong correlation effects and deviations from simple itinerant 5\textit{f}$^3$ band theory \cite{fujimori2019}.

The combined evidence from our thin-film valence spectra and the PES data collected on single crystals of UTe$_2$ by Miao et al. \cite{Miao}, Sundermann et al. \cite{Sundermann2025UTe} and Fujimori et al. \cite{fujimori2019} indeed indicates that electronic structure of UTe$_2$ shows intermediate-valence character, where both 5\textit{f}$^2$ and 5\textit{f}$^3$ configurations contribute to the ground state. In this picture, the spectral weight at higher binding energies primarily originates from incoherent 5\textit{f}$^2$ multiplet states, while the near-$E_{\mathrm{F}}$ part of the spectrum reflects hybridized 5$f$-derived states that form due to interaction with conduction electrons.

When comparing valence band XPS data collected with the He II UPS data (Fig. 3) (higher energy resolution and greater surface sensitivity in UPS), both techniques consistently show the metallic Fermi-edge cutoff within their respective energy resolutions. However, UTe$_2$ is known to accommodate various types of defects and local structural deviations \cite{svanidze,aoki2024}, affecting superconductivity and low-energy spectroscopy features. Such imperfections together with the finite energy resolution of laboratory-based XPS, surface or termination effects, as well as thin-film-related factors such as strain and slight deviations from stoichiometry, can all contribute to the suppression or broadening of coherent 5$f$ spectral weight near $E_{\mathrm{F}}$.

As the stoichiometry further increases to Te:U = 3, we observe a continued decrease of the 5\textit{f} emission. The maximum dominating the valence band becomes more asymmetric, with the spectral weight moving more towards $E_{\mathrm{F}}$. The Fermi level cutoff is not observed, the weak residual intensity at BE = 0 is consistent with a strongly suppressed DOS at $E_{\mathrm{F}}$ (pseudogap or gap-like behavior). This is compatible with the $\alpha$-UTe$_3$ phase, known to exhibit semiconducting behavior \cite{BLAISE1981417}. On the other hand, the $\beta$-UTe$_3$ allotrope is metallic, with the Sommerfeld coefficient enhanced to 130 mJ/mol·K$^2$ \cite{thomas2025enhancedtwodimensionalferromagnetismvan}. The situation with the highest available Te:U content of 4.8 appears qualitatively similar to the case of Te:U = 3. This ratio lies close to the 1:5 region reported in the U–Te phase space; the lower uranium fraction naturally weakens the  U~4\textit{f} signal, so XPS-based stoichiometry here carries somewhat larger relative uncertainty. As such, while the trend in spectral features remains informative, the precise Te:U ratio at this composition should be interpreted with caution and ideally confirmed by complementary methods.

We can conclude that the spectroscopic data revealed pronounced composition-dependent changes in the position and spectral distribution of the 5\textit{f} states with respect to the Fermi level. In the next section, we use \textit{ab initio} calculations to rationalize these trends in terms of stoichiometry- and structure-dependent bonding within the U–Te series. The results are in the Section IV below.

\section{\label{sec:level1}Calculations and discussion}

The relatively simple scalar relativistic DFT calculations cannot capture the complex many-body effects important for the 5\textit{f} states and their (de)localization. For a quantitative description, DFT+\textit{U} or DFT+DMFT techniques would be needed. Those would, however, require an individual approach for each compound (as e.g. different values of Coulomb \textit{U}) and the trends of bonding variations would be blurred. As we prefer comparison within the same framework, we used the FPLO and (L)APW + lo methods with scalar relativistic approximations, which are computationally feasible even for large supercells. Such an approach may provide a good account of cohesion properties. We cannot expect agreement with the size of magnetic moments (affected by orbital moments and spin-orbit coupling). Naturally also details of the spectral function determining the valence-band spectra are affected by e-e correlations if the correlation energies are comparable with the 5\textit{f} line width. Hence the relation of DFT DOS and experimental spectra is merely indirect (cf. Fig. 6). This is particularly true for $\alpha$-UTe$_3$, in which the insulating ground state arises due correlation effects.  
Fig. 6 summarizes the total and partial U-5$f$,6$d$ and Te-5$p$ DOS for all the known binary compounds. All the cases exhibit the 5$f$ states exchange-split (by 1-2 eV) appearing in the LSDA approximation even for UTe$_2$ with a non-magnetic ground state. (For consistency with other binaries, spin-polarized UTe$_2$ was taken despite its known non-magnetic ground state; the presence of sizeable effective moments in the experimental susceptibility indicates an underlying spin splitting \cite{ran}.)

Table I summarizes the compositions and their corresponding magnetic and electronic properties, which serve as key reference points for our study. To place the U–Te series into a broader actinide context, it is useful to recall that for uranium compounds, the Hill limit provides a primary guideline about the degree of 5\textit{f} delocalization being modulated by the U-U spacing. The Hill picture suggests that when U-U distances are below $\sim3.6$ \AA, there is considerable overlap of 5\textit{f} orbitals, which can result in electrons participating in metal bonding and, e.g., electrical conduction. For greater U-U separations, which applies for all the U-Te compounds listed here, itinerant 5\textit{f} behaviour may be gradually lost, unless other delocalizing mechanisms (5\textit{f}-ligand hybridization) overtake the dominant role. Nevertheless, 5\textit{f} electrons in uranium compounds typically retain a degree of participation in bonding and conduction \cite{sech}. This nuanced behavior extends to the U-Te series, where our observations demonstrate a variety of valence band characteristics (Fig. 3), shaped by significant hybridization effects and strong electron correlations.

\begin{figure}[H]
    \centering
    \begin{subfigure}{0.45\linewidth}
        \centering
        \includegraphics[width=\linewidth]{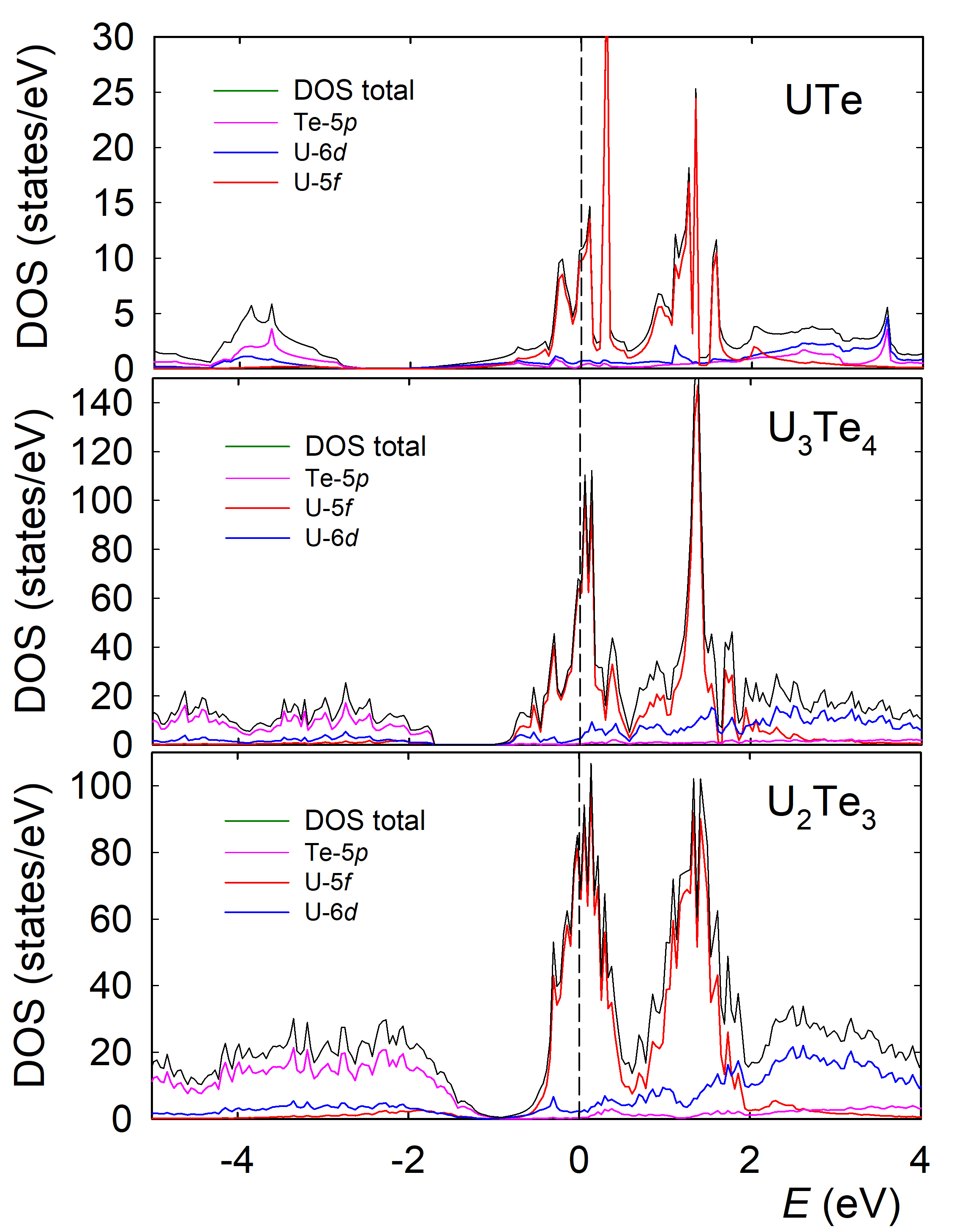}
        \caption{}
    \end{subfigure}
    \hfill
    \begin{subfigure}{0.45\linewidth}
        \centering
        \includegraphics[width=\linewidth]{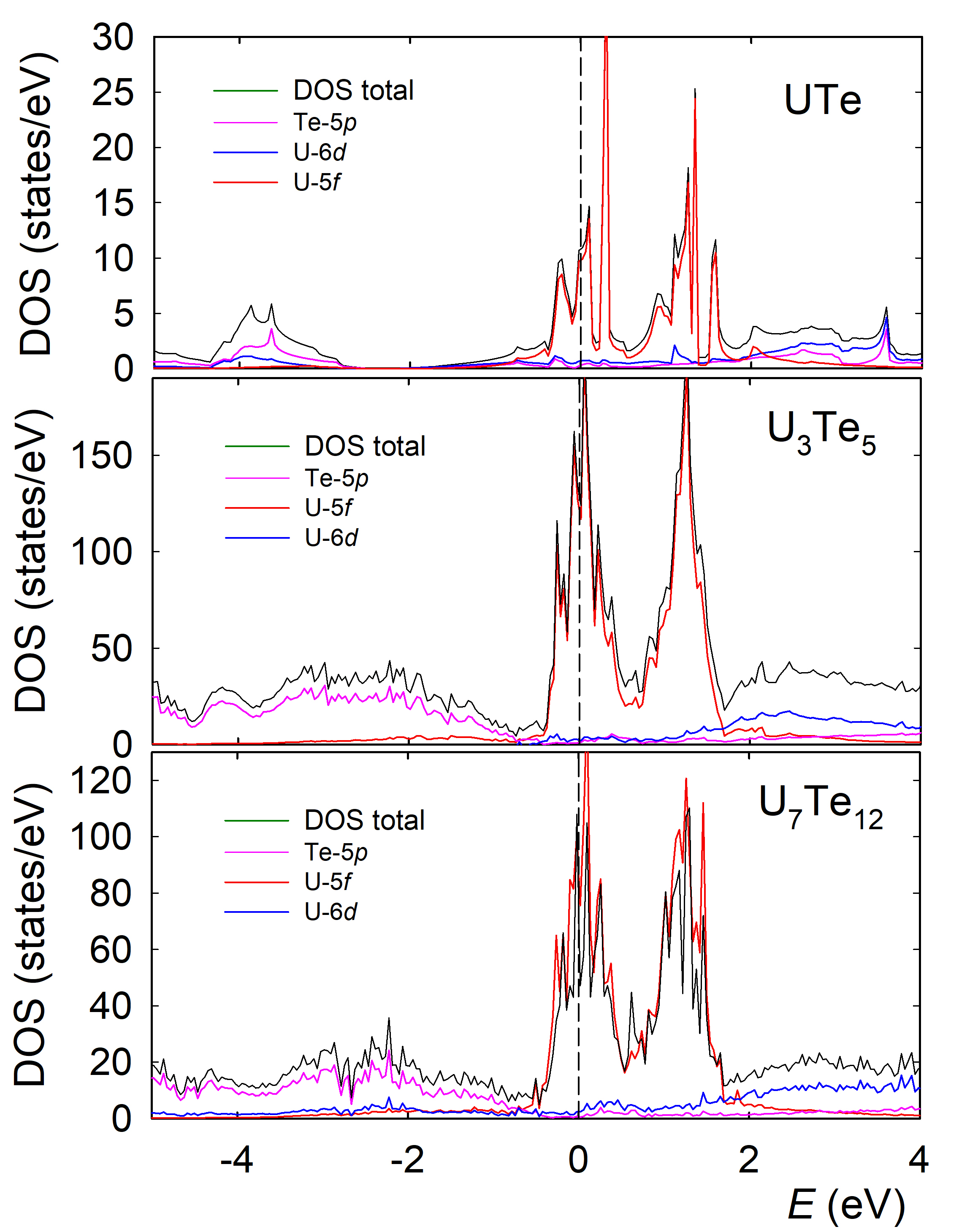}
        \caption{}
    \end{subfigure}

    \medskip

    \begin{subfigure}{0.45\linewidth}
        \centering
        \includegraphics[width=\linewidth]{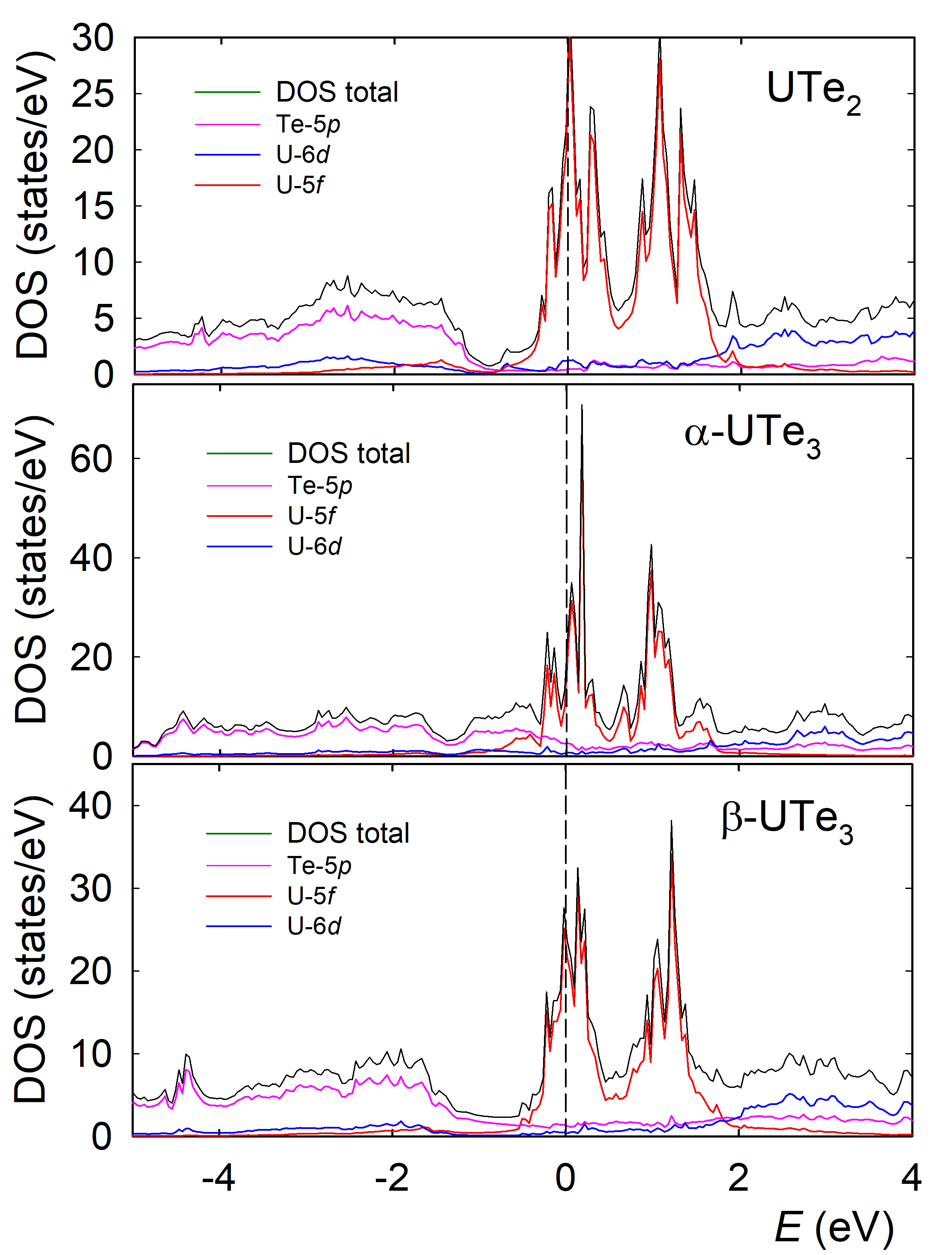}
        \caption{}
    \end{subfigure}
    \hfill
    \begin{subfigure}{0.45\linewidth}
        \centering
        \includegraphics[width=\linewidth]{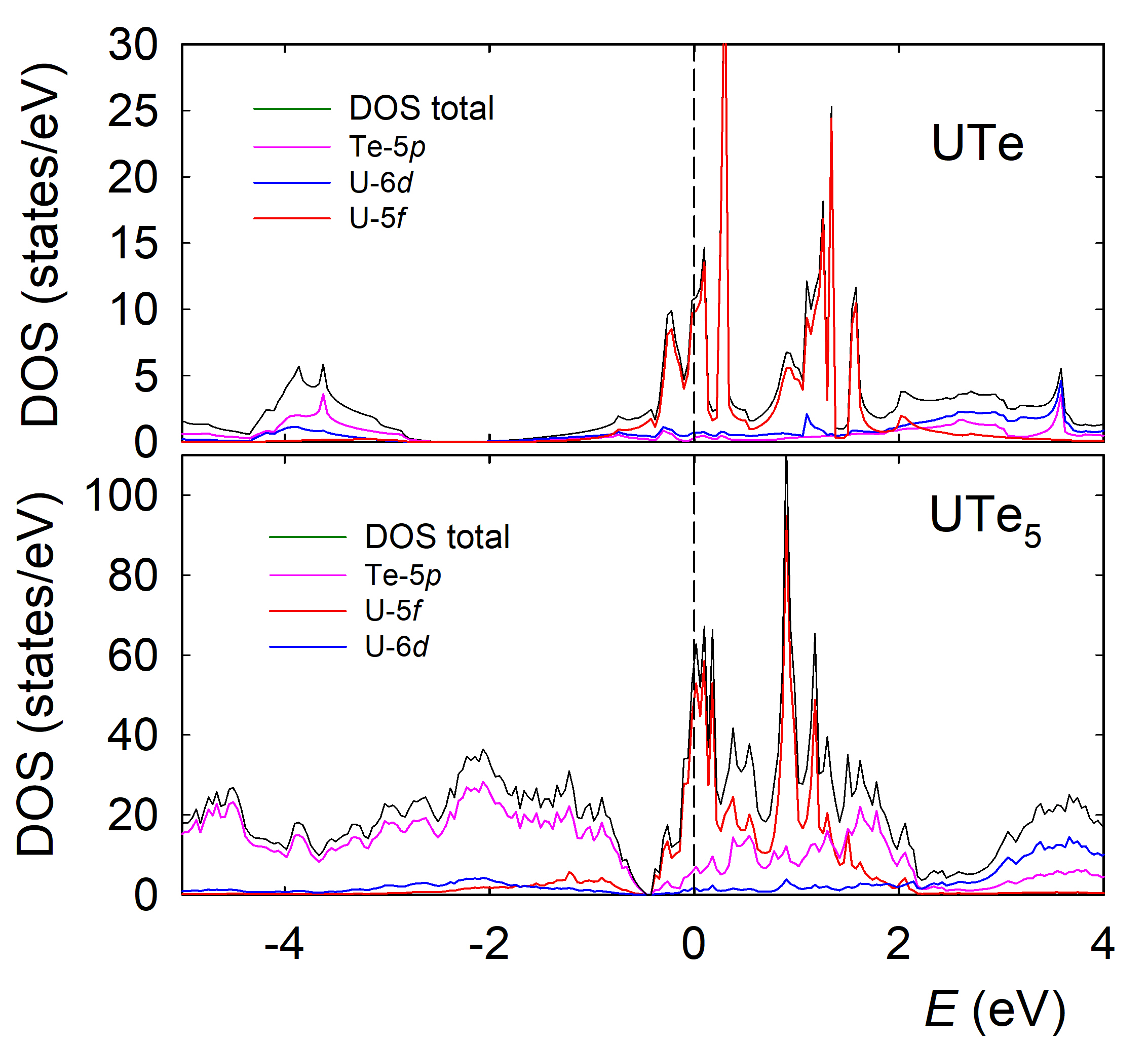}
        \caption{}
    \end{subfigure}

    \caption{(a)–(d) Calculated total and partial densities of states (DOS) for various U-Te bulk compositions. The total DOS (black) is decomposed into U-5$f$ (red), U-6$d$ (blue), and Te-5$p$ (magenta) contributions. The vertical dashed line marks the Fermi level ($E_{\mathrm{F}}$ = 0 eV). The prominent features of each spectrum are the exchange-split 5$f$ states. One should notice different scales of DOS, depending on the size of the unit cell.}
    \label{fig:fourpanel}
\end{figure}

Changes in bonding between different compounds are reflected in variable occupancies of particular electronic states. Individual occupancies sorted by the compounds and different types of positions, as obtained from the FPLO calculations, are shown in Table II. The population analysis of FPLO allows to distinguish between the smaller net electron charge and the gross charge, which includes area of wavefunctions overlap between ligands. Naturally a compound with several non-equivalent crystal structure sites and elements can have somewhat different population of states for each site.

Table II. Calculated occupancies of different types of valence states of various uranium and tellurium sites corresponding to the crystal structures listed in Table I. Note that for each site we display the net charge (upper line), while the gross charge is shown in the lower line.
\begin{ruledtabular}
\begin{tabular}{ccccccc}
 &At. site &Q(7\textit{s}) &Q(6\textit{d}) &Q(5\textit{f}) &Q(5\textit{s}) &Q(5\textit{p})\\
\hline
 
UTe &\makecell{U(4a)\\ \\Te(4b)} &\makecell{0.11 \\ 0.31\\ \\ \\} &\makecell{1.16 \\ 2.04\\ \\ \\} &\makecell{2.86 \\ 3.10\\ \\ \\} &\makecell{ \\ \\ 1.65\\1.76 \\} &\makecell{ \\ \\ 3.47 \\4.12} \\
\hline

U$_3$Te$_4$ &\makecell{U1(12a)\\ \\ Te (16c)} & \makecell{0.13\\0.35 \\ \\ \\} &\makecell{0.88\\1.66\\ \\ \\} &\makecell{2.97\\3.17\\ \\ \\} &\makecell{ \\ \\ 1.82\\1.86 \\} &\makecell{ \\ \\ 3.84 \\4.37}\\
\hline

U$_2$Te$_3$ &\makecell{U1 (4c)\\ \\ U2 (4c) \\ \\ Te1 (4c)\\ \\Te2 (4c)\\ \\Te3 (4c)} &\makecell{0.15 \\0.39\\0.15\\0.39\\ \\ \\ \\ \\ \\ \\}
 & \makecell{0.90\\1.73\\0.84\\1.62 \\ \\ \\ \\ \\ \\ \\} &\makecell{2.88\\3.09\\2.91\\3.10 \\ \\ \\ \\ \\ \\ \\ } &\makecell{\\ \\ \\ \\ 1.78\\ 1.85\\ 1.76\\ 1.84\\ 1.82\\ 1.88} &\makecell{\\ \\ \\ \\ 3.74\\ 4.35\\  3.72\\ 4.30\\ 3.80\\4.37}\\
\hline

U$_3$Te$_5$ &\makecell{U1(4c) \\ \\ U2(4c) \\ \\ U3(4c) \\ \\Te1 (4c)\\ \\ Te2 (4c) \\ \\ Te3 (4c) \\ \\Te4 (4c) \\ \\ Te5 (4c)} & \makecell{0.15 \\ 0.38 \\ 0.14 \\ 0.38 \\0.14 \\ 0.38 \\ \\ \\ \\ \\ \\ \\ \\ \\ \\ }
 & \makecell{0.88 \\ 1.72 \\0.94 \\1.83 \\0.90 \\1.75\\ \\ \\ \\ \\ \\ \\ \\ \\ \\}
 & \makecell{2.81 \\ 3.03 \\ 2.74 \\2.99 \\2.79 \\3.02\\ \\ \\ \\ \\ \\ \\ \\  \\ \\} & \makecell{ \\  \\  \\ \\ \\ \\1.82\\1.87\\1.76\\1.82\\1.82\\1.87\\1.78\\1.85\\1.82\\1.86} & \makecell{\\ \\ \\ \\ \\ \\3.81\\4.37\\3.79\\4.24\\3.80\\4.34\\3.79\\4.35\\3.76\\4.30} \\
 \hline

U$_7$Te$_{12}$ & \makecell{U1(1a) \\ \\ U2(3k)\\ \\U3(3j) \\ \\ Te1 (3k) \\ \\ Te2 (3k)\\ \\ Te3 (3j)\\ \\ Te4 (3j)} & \makecell{0.16\\0.39\\0.16\\0.40\\0.15\\0.39\\ \\ \\ \\ \\ \\ \\ \\ \\}
 & \makecell{0.85 \\1.67\\0.95\\1.84\\0.93\\1.81\\ \\ \\ \\ \\ \\ \\ \\ \\} & \makecell{2.79\\3.01\\ 2.70\\2.96\\2.74\\2.98\\ \\ \\ \\ \\ \\ \\ \\ \\ } & \makecell{ \\  \\  \\ \\ \\ \\1.87\\1.89\\1.83\\1.87\\1.79\\1.85\\1.79\\1.84} &\makecell{ \\  \\  \\ \\ \\ \\3.83\\4.34\\3.78\\4.32\\3.76\\4.32\\3.79\\4.25}\\
\hline

UTe$_2$ &\makecell{U1(4i) \\ \\ Te1 (4j)\\ \\ Te2 (4h)}
& \makecell{0.12\\0.36 \\ \\ \\ \\ \\} &\makecell{0.97\\1.89\\ \\ \\ \\ \\} &\makecell{2.69\\2.96 \\ \\ \\ \\ \\} &\makecell{\\ \\ 1.77\\ 1.85\\ 1.91\\1.89} &\makecell{\\ \\ 3.76\\ 4.32\\ 3.63\\4.14}\\
\hline

U$_2$Te$_5$ &\makecell{U1(4i)\\ \\U2(4i)\\ \\ Te1 (4i) \\ \\ Te2 (4i) \\ \\ Te3 (4i)\\ \\ Te4 (4i) \\ \\Te5 (4i)} &\makecell{0.13\\0.38\\0.13\\0.37 \\ \\ \\ \\ \\ \\ \\ \\ \\ \\} &\makecell{1.03\\2.00\\1.04\\1.99\\ \\ \\ \\ \\ \\ \\ \\ \\ \\} &\makecell{2.56\\2.88\\2.59\\2.90\\ \\ \\ \\ \\ \\ \\ \\ \\ \\} &\makecell{\\ \\ \\ \\ 1.77\\ 1.85\\1.76 \\ 1.84\\1.92 \\1.89 \\ 1.97 \\ 1.92\\1.97\\1.92} &\makecell{\\ \\ \\ \\ 3.71\\ 4.27\\3.64 \\ 4.22\\3.60 \\4.12 \\ 3.54 \\ 4.05\\3.54\\4.04}\\
\hline
$\alpha$-UTe$_3$&\makecell{U1(2e) \\ \\ Te1(2e) \\ \\ Te2(2e) \\ \\ Te3(2e)} &\makecell{0.18\\0.43\\ \\ \\ \\ \\ \\ \\ } &\makecell{0.98\\1.83\\ \\ \\ \\ \\ \\ \\ } &\makecell{2.68\\2.95\\ \\ \\ \\ \\ \\ \\ } &\makecell{\\ \\1.89\\1.88\\1.99\\1.91\\2.08\\1.95\\ } &\makecell{\\ \\3.65\\4.14\\3.59\\4.09\\3.69\\4.07\\}\\
\hline

$\beta$-UTe$_3$&\makecell{U1(4c) \\ \\ Te1(4c) \\ \\ Te2(4c) \\ \\ Te3(4c)} &\makecell{0.13\\0.36\\ \\ \\ \\ \\ \\ \\ } &\makecell{0.91\\1.82\\ \\ \\ \\ \\ \\ \\ } &\makecell{2.74\\3.01\\ \\ \\ \\ \\ \\ \\ } &\makecell{\\ \\2.10\\1.94\\2.10\\1.94\\1.79\\1.85\\ } &\makecell{\\ \\3.53\\3.99\\3.52\\3.98\\3.80\\4.36\\}\\
\hline

UTe$_5$ &\makecell{U1(4c) \\ \\ Te1(8c) \\ \\ Te2(4c) \\ \\ Te3(4c) \\ \\Te4(4a)} &\makecell{0.13\\0.38 \\ \\ \\ \\ \\ \\ \\ \\} &\makecell{1.05\\2.07 \\ \\ \\ \\ \\ \\ \\ \\} &\makecell{2.60\\2.91\\ \\ \\ \\ \\ \\ \\ \\} &\makecell{\\ \\ 1.97\\ 1.91\\ 1.97\\ 1.92\\ 2.05\\ 1.95\\ 1.98\\} &\makecell{\\ \\ 3.55\\ 4.04\\ 3.60\\ 4.08\\3.40 \\ 3.87\\ 3.49\\}\\
\end{tabular}
\end{ruledtabular}

%{Our calculations reveal at larger U-U distances, the 5\textit{f}-5\textit{f} interactions become more distinct, and hybridization with Te p states is less pronounced (Table II). Conversely, as U-U distances decrease, hybridization with Te p states becomes significantly stronger, enhancing interactions between U-5\textit{f} electrons and the ligand orbitals. The bonding complexity is further compounded by various Te-Te configurations that manifest across different stoichiometries. For instance, in UTe$_2$, Te-Te interactions form one-dimensional chains, complicating a simple (U$^{4+}$)(Te$^{2-}$)2 representation. Instead, the presence of intermediate Te-Te bond lengths—shorter than the van der Waals distance yet longer than a full Te-Te single bond—suggests a mixed oxidation state for tellurium, resulting in a non-integral formal charge \cite{christovam}.}

Considering the polar character of bonding between more electronegative Te and less electronegative U, it is interesting to explore how the possible charge transfer affects the occupancy of individual types of states. The difference in electronegativities between U and Te implies that the U states should be somewhat depleted especially for higher Te concentrations. The enhancement of occupancies of Te states complements the picture. 

For clarity, the main trends extracted from Table II are visualized in Fig. 7, which summarizes the evolution of the occupancy of the total U-valence states, the 5\textit{f} contribution, and the non-\textit{f} (mainly 6\textit{d}) contribution across the U–Te series. One can notice the expected decreasing tendency of the total U valence occupancy with increasing Te concentration, which saturates above the 1:2 composition. The 5\textit{f} occupancy is highest for the 3:4 phase, where the non-\textit{f} occupancy exhibits a pronounced dip. The figure also compares spin-polarized (LSDA) and non-magnetic (LDA) calculations for UTe$_2$. The difference remains small, but cannot be considered negligible.

\begin{figure}[h!]
    \centering

    \begin{subfigure}{0.32\textwidth}
        \centering
        \includegraphics[width=\linewidth]{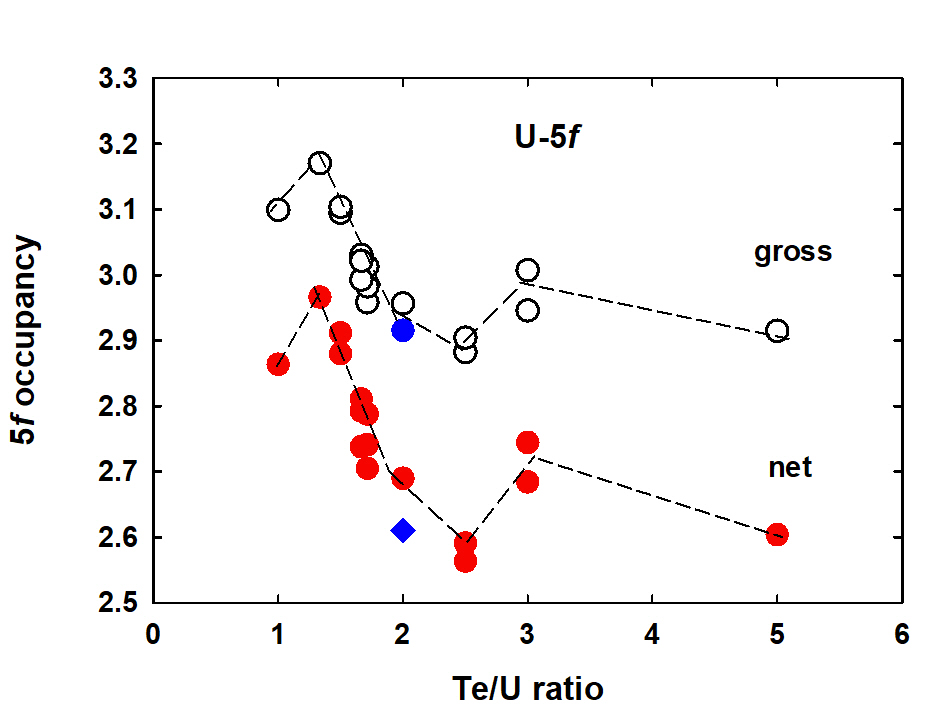}
        \caption{}
    \end{subfigure}
    \hfill
    \begin{subfigure}{0.32\textwidth}
        \centering
        \includegraphics[width=\linewidth]{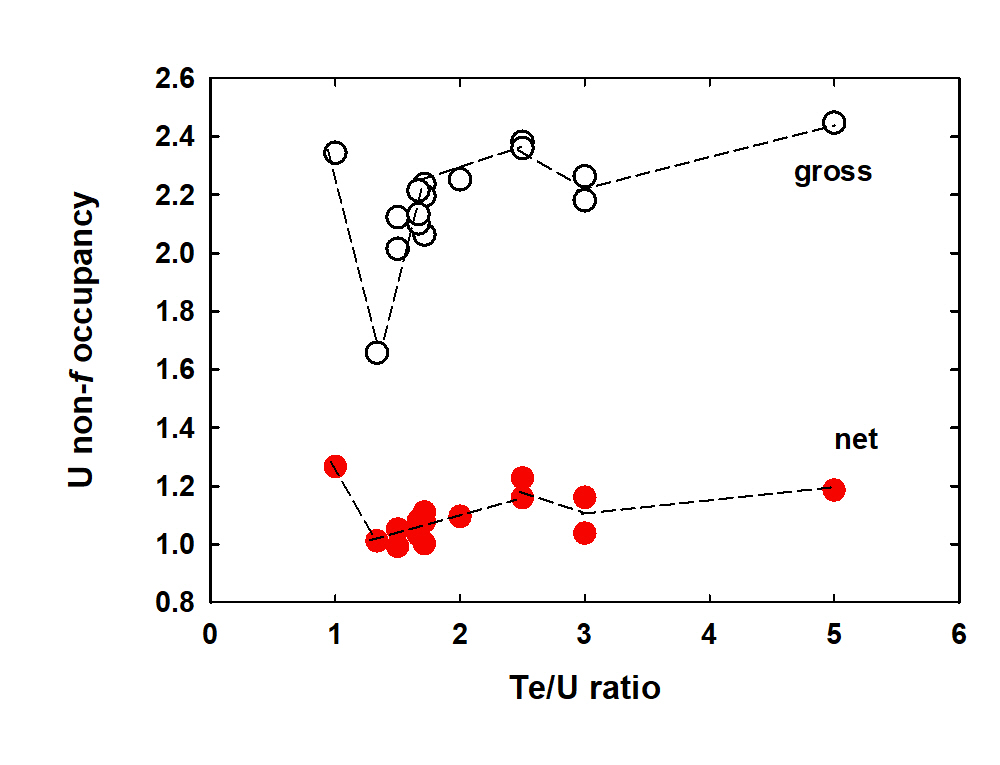}
        \caption{}
    \end{subfigure}
    \hfill
    \begin{subfigure}{0.32\textwidth}
        \centering
        \includegraphics[width=\linewidth]{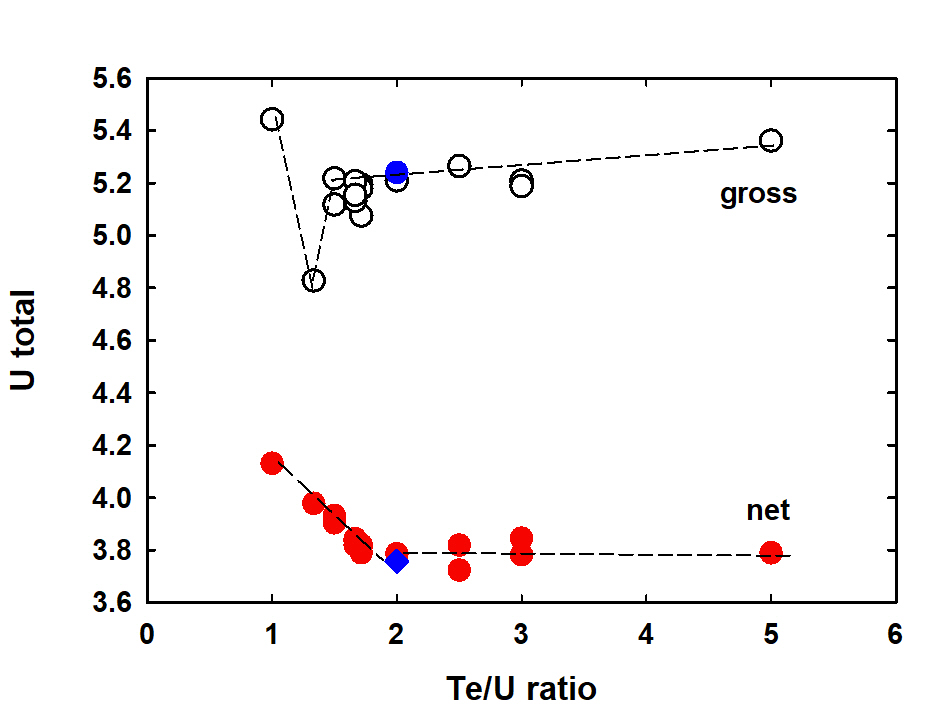}
        \caption{}
    \end{subfigure}

    \caption{Evolution of calculated uranium valence occupancies across the U--Te series, presented as net and gross FPLO populations: (a) U-5$f$ occupancy, (b) non-\textit{f} occupancy (primarily U-6$d$), and (c) total U-valence states occupancy (7$s$+6$d$+5$f$). Multiple points for a given composition reflect inequivalent U sites in the crystal structure (and two polymorphs in the case of UTe$_3$). For UTe$_2$, results of spin-polarized (LSDA) and non-magnetic (LDA) calculations are compared.}
    \label{fig:7}
\end{figure}

Inspection of Table II and Fig. 7 reveals several notable trends.
a) The 5\textit{f} occupancies are particularly high in UTe and U$_3$Te$_4$. The net 5\textit{f} occupancies are only slightly lower than 3.0, the gross occupancies even exceed 3.0. Such  numbers are exceptional among U compounds, comparable perhaps with UCd$_{11}$ with \textit{n}$_{5f}$ considered close to 3 on the basis of X-ray absorption experiment \cite{Nasreen}. Between the two, the higher occupancy was obtained in U$_3$Te$_4$, which has, however, a lower 6\textit{d} occupancy. The total net U occupancy (including small 7\textit{s} occupancies not shown in Table II) is the highest in UTe, reaching 4.13, while it is 3.98 in U$_3$Te$_4$. The relatively lower difference between net and gross occupancies at the 5\textit{f} states comparing to the 6\textit{d} states, which appears for all compositions, reflects the larger 6\textit{d} spatial extent and stronger involvement in bonding. 

b) On the opposite end, low 5\textit{f} occupancies around 2.6 were found as expected for UTe$_5$, but also for UTe$_2$ and U$_2$Te$_5$. This draws a clear division line between UTe$_2$ and U$_7$Te$_{12}$, the latter having higher average net 5\textit{f} occupancy despite not so different stoichiometry. 

c) With respect to Te occupancies, UTe is the outstanding case with low total (5\textit{s}+5\textit{p}) occupancy (gross value 5.88), inferior to the nominal free-ion value of 6 (5\textit{s}$^2$, 5\textit{p}$^4$). We can therefore conclude that UTe exhibits the least polar bonding in the series. In UTe$_5$, the gross average Te occupancy is only moderately higher, which is largely a trivial consequence of the 1:5 stoichiometry, due to which the polar bonding affects more U than Te. Consistently, the net Te occupancies in UTe$_5$ are substantially higher than in UTe, indicating that U–Te hybridization has a weaker impact on Te-derived states. 

d) UTe$_2$ exhibits a pronounced difference between the two Te sites, one with the gross total occupancy 6.01, the other one with 6.14. The difference of the two Te sites, deduced in \cite{christovam} from fully relativistic calculations, appears already on the level of scalar relativistic GGA calculations. 

A natural question is whether UTe$_2$ is anomalous in its bonding compared to other U-Te compounds. The answer appears to be yes. One of the two Te sites (Te2 - 4\textit{h}) is, due to the specific crystal structure, appearing comparatively inert as to the bonding with U. The remaining U-Te bonding network within UTe$_2$, however, is markedly more polar than in the other binary U–Te compounds, with the main consequence being a reduced occupancy of U-5\textit{f} (and other) states. The analysis in \cite{christovam}, based on comparison of UTe$_2$ with UO$_2$, suggests that Te2 donates electron density primarily into U 6\textit{d}-derived states.

We stress that the numerical results should be interpreted as indicative of tendencies only. The omission of spin-orbit coupling and especially of the e-e correlations beyond DFT (Hubbard \textit{U}) introduce a bias against formation of a band gap, which can affect the occupancies and 5\textit{f} localization if such a gap is formed in reality. The computations cannot treat all such possibilities on the same footing and compare, for example, the total energies. Hence we cannot contribute to the discussion if the 5\textit{f} states in UTe$_2$ are related to localized 5\textit{f}$^2$, 5\textit{f}$^3$, or itinerant 5\textit{f} states with average occupancy between 2 and 3. 
Nevertheless, we observe a consistent trend toward reduced 5\textit{f} occupation comparing to U metal and U-Te compounds with lower (or comparable) Te concentration. The spectroscopic data reveal that the U-4\textit{f}$_{7/2}$ peak is shifted towards higher binding energies, by 0.8 eV with respect to U metal, or by 0.5 eV with respect to UTe (Figs. 1(left) and 2). This shift can be related to depletion of the U-derived valence-band states due to bonding with Te. The asymmetric shape of the peaks indicates that the 4\textit{f} hole is screened by the 5\textit{f} states, i.e., the 5\textit{f} states cannot be fully localized. One can compare this shift with the 5\textit{f}$^2$ localized case in UPd$_3$, in which the 5\textit{f} screening channel is forbidden and the less efficient screening by non-\textit{f} valence electrons shifts the main peak by 2 eV towards higher binding energies \cite{fujimori2012}.

Valence-band spectra acquired in both XPS and UPS modes further show that in UTe$_2$ 5\textit{f} shifted away from $E{\mathrm{F}}$ into a broad maximum at finite binding energy, while a metallic Fermi-edge cutoff remains visible (Fig. 3). In particular, the He II spectrum exhibits a broad maximum around 0.5 eV below $E_{\mathrm{F}}$ in the UTe$_2$ film and the near-$E_{\mathrm{F}}$ spectral weight is considerably reduced compared to UTe. The He I spectrum, with a low photoexcitation cross section of the 5\textit{f} states, also shows a distinct maximum near 0.5 eV (Fig. 4), which can be attributed, at least partly, to the 6\textit{d} emission in agreement with Ref. \cite{christovam}. By comparing the He II and He I spectra for UTe and UTe$_{1.3}$, the latter likely corresponding to U$_3$Te$_4$ (Appendix~A, Fig. A2), we deduce that the 5\textit{f} states contribute to the maxima at 0.7-0.8 eV more than to the emission at $E_{\mathrm{F}}$. For Te concentrations above 2, the valence-band maximum shifts slightly back towards $E_{\mathrm{F}}$, consistent with a trend of the shift of 4\textit{f} peaks towards lower binding energies (Fig. 2). 

In addition, the U-6\textit{p}$_{3/2}$ spectra were collected for selected Te concentrations (Fig. 5). A shift of several eV relative to metallic U is observed, which further increases with increasing Te concentration. A similarly strong shift has been reported for uranium hydrides, where the U 6\textit{p}$_{3/2}$ peak shifts to approximately 18 eV binding energy \cite{HAVELAU-H3}. The much larger shift compared to the 4\textit{f} lines can be qualitatively understood by the large spatial extent of the U-6\textit{p} lines (comparing to the 4\textit{f} states located deep inside the core), and their sensitivity to the evolution of the 6\textit{d} states hybridizing with the states of electronegative ligands \cite{Havela_2023}.

More broadly, among uranium compounds formed with different $p$-block ligands, uranium tellurides stand out by the pronounced sensitivity of their electronic structure to stoichiometry and local coordination. This behavior reflects a strong and composition-dependent hybridization between U-5$f$ and Te-$p$ states, which directly reshapes the electronic structure in the vicinity of the Fermi level. By contrast, uranium compounds with lighter $p$-block ligands such as Si~\cite{sarma}, Ga~\cite{reihl1985,GOUDER20017}, and Ge~\cite{AlexU-Ge} form comparatively rigid bonding frameworks, in which changes in stoichiometry primarily modulate the overall spectral weight while leaving the valence-band line shape largely unchanged. This contrast highlights the unique role of tellurium in enabling a hybridization-driven reconstruction of the U-5$f$ manifold that is absent in more rigid U–$p$ systems.

\section{\label{sec:level1}Conclusions}
We have presented a systematic photoelectron spectroscopy study of uranium telluride thin films spanning the compositional range of the U–Te phase diagram. By employing a thin-film approach, we achieved continuous control over stoichiometry and accessed compositions that are difficult to prepare and investigate in bulk form. XPS and UPS effectively capture variations in bonding and metallicity across different Te:U compositions by identifying $f$-derived features within the valence bands. The experimental trends are supported by \textit{ab initio} calculations performed on the same footing for all compositions, which provide insight into charge redistribution and its evolution across the U–Te series. Taken together, these results highlight uranium tellurides as a materials family in which hybridization effects, shaped by ligand chemistry and local structure, provide a useful organizing perspective for understanding their diverse electronic ground states.

\begin{acknowledgments}
We acknowledge the support of Czech Science Foundation under the grant no. 22-19416S. The samples were prepared in the framework of the EARL project of the European Commission Joint Research Centre, ITU Karlsruhe. Physical properties measurements were performed in the Materials Growth and Measurement Laboratory (http://mgml.eu/) supported within the program of Czech Research Infrastructures (Project No. LM2023065). S.G.A. acknowledges support from the Grant Agency of Charles University (GAUK, project no. 416925, starting in 2025), which supports his ongoing contributions to data analysis. We thank Shin-ichi Fujimori for providing the Te-4\textit{d} spectra of UTe$_2$, presented in his seminar “Electronic structures of uranium compounds studied by photoelectron spectroscopy” (Charles University, May 2021), and reproduced with permission in the Appendix~A, Fig. A4.
\end{acknowledgments}

\appendix
\section{Supplementary figures}

\setcounter{figure}{0}
\renewcommand{\thefigure}{A\arabic{figure}}

\begin{figure}[h!]
\centering
\includegraphics[width=0.7\textwidth]{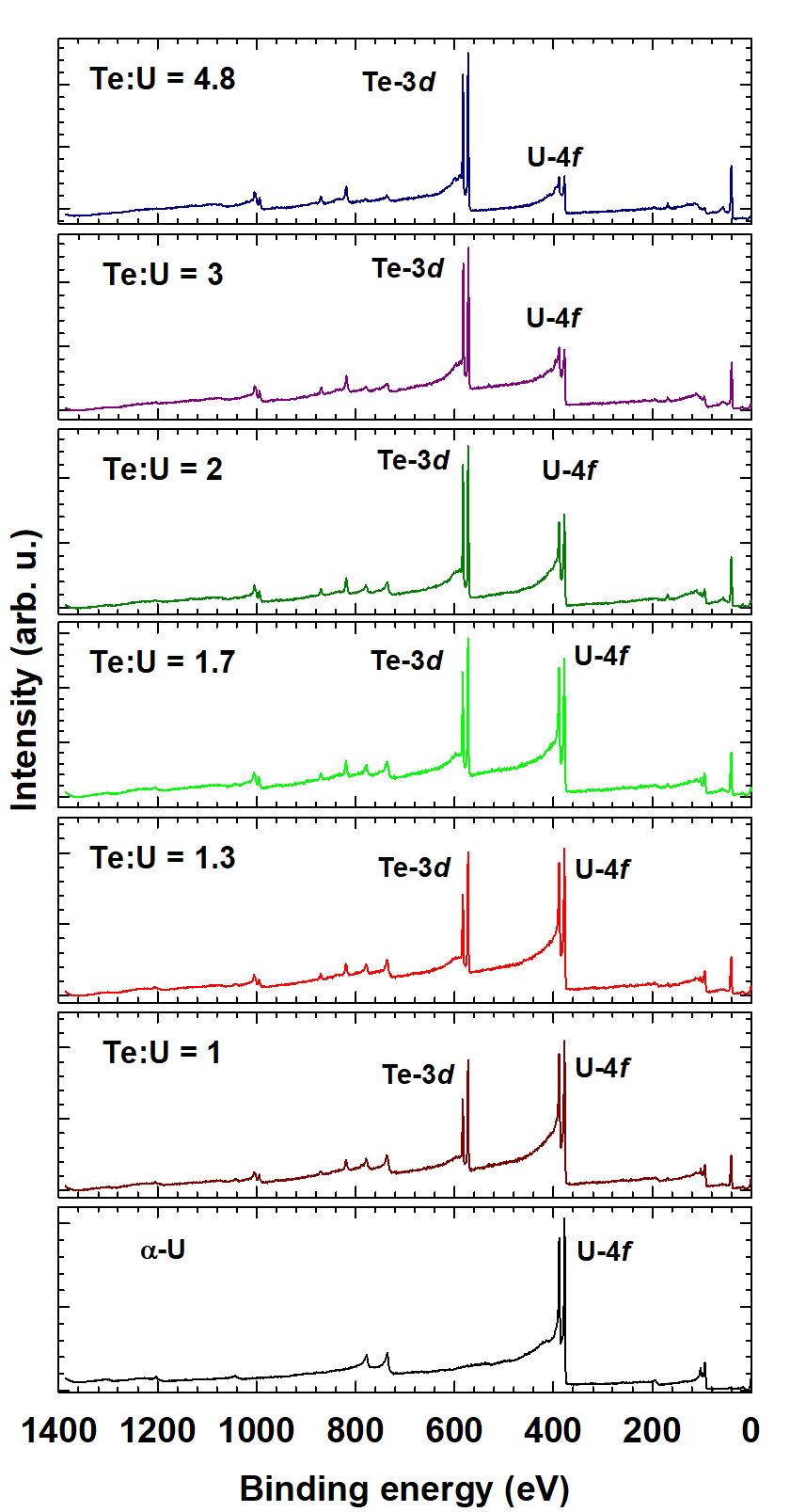}\,
\caption{XPS spectra of various U-Te thin films, obtained with Al-K$_\alpha$ radiation, demonstrating the absence of O and C contaminants. The spectra highlight the increasing Te:U intensity ratio with higher Te content.}
\label{fig:A1}
\end{figure}

\begin{figure}[h!]
\centering
\includegraphics[width=0.7\textwidth]{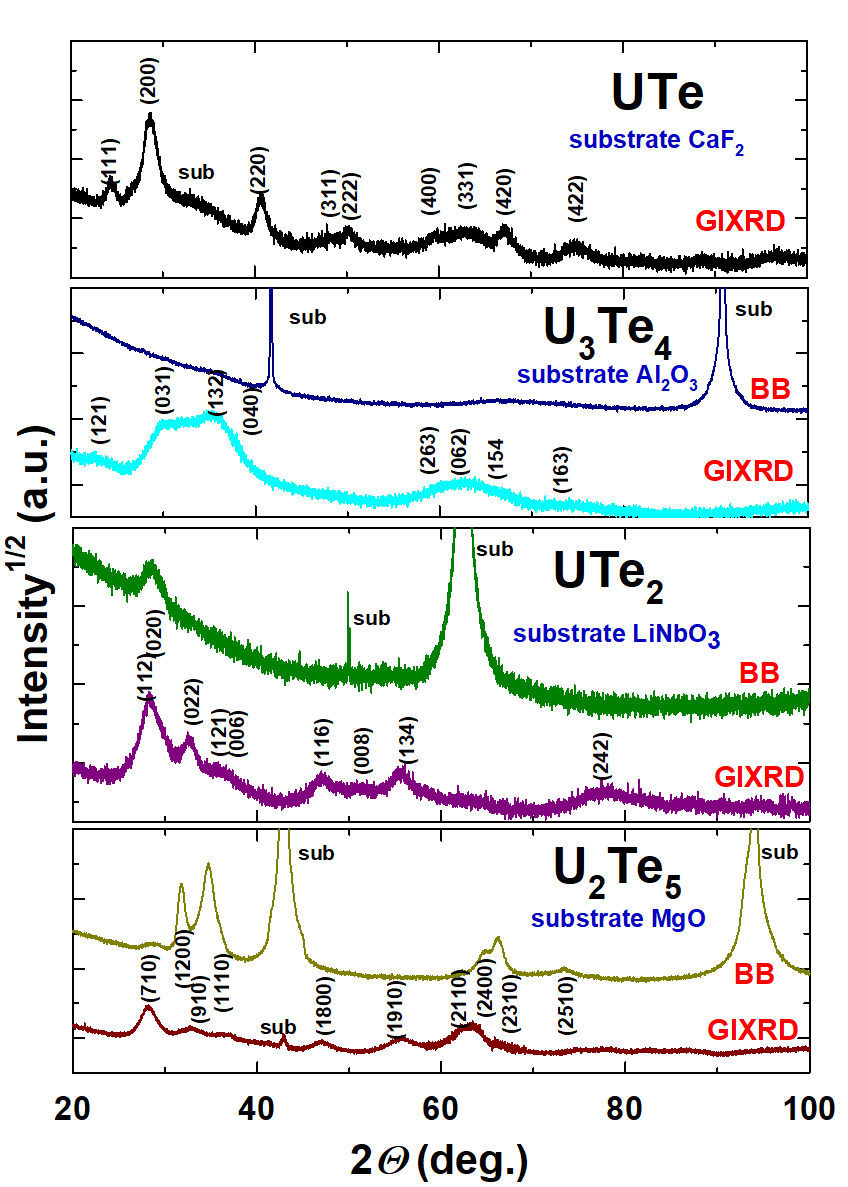}\,
\caption{Room-temperature XRD patterns of thin films of selected U-Te compositions obtained in Bragg-Brentano (BB) geormetry and under grazing incidence (GIXRD) with a constant angle of incidence of 1.5$^{\circ}$ and Cu-K$\alpha$ radiation ($\lambda$ = 0.15418 nm). The weak diffraction features reflect the limited diffracting volume of the thin (tens of nanometres) films. The UTe$_2$ film exhibits a strong b-axis preferential out-of-plane orientation (fiber texture).}
\label{fig:A2}
\end{figure}

\begin{figure}[h!]
\centering
\includegraphics[width=0.7\textwidth]{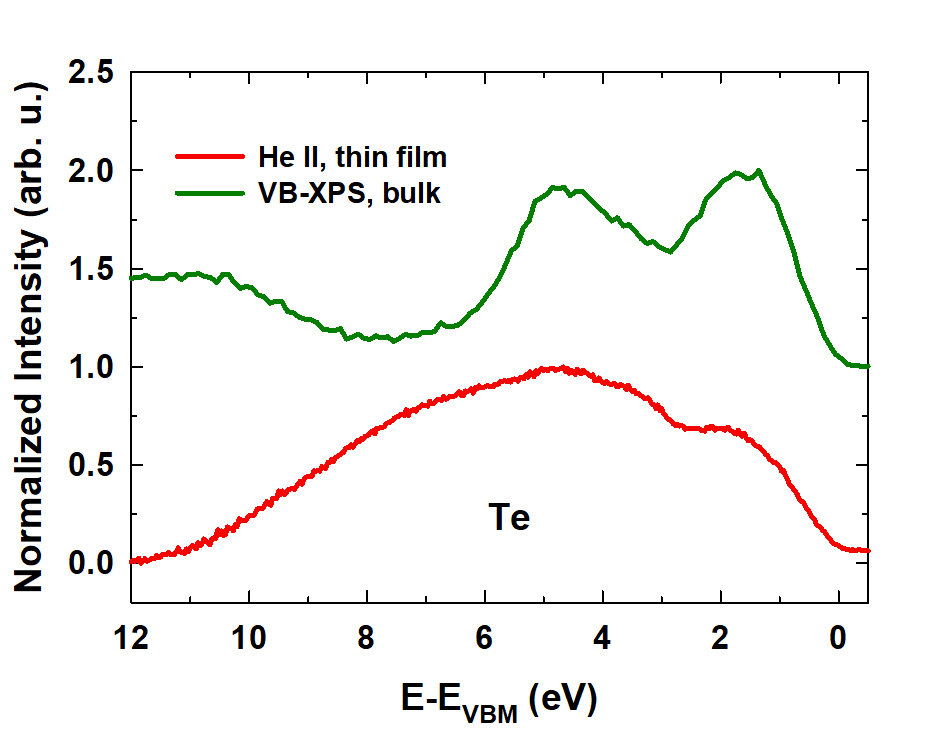}\,
\caption{Valence band photoemission spectra measured using He II UPS and Al K$\alpha$ XPS for the Te film and sputter-cleaned bulk Te samples, respectively. The energy scale is calibrated to the position of the valence band maximum (VBM). Differences in spectral line shape and peak intensity arise from the different photoemission cross-sections for He II and Al K$\alpha$ excitation.}
\label{fig:A3}
\end{figure}

\begin{figure}[h!]
\centering
\includegraphics[width=0.7\textwidth]{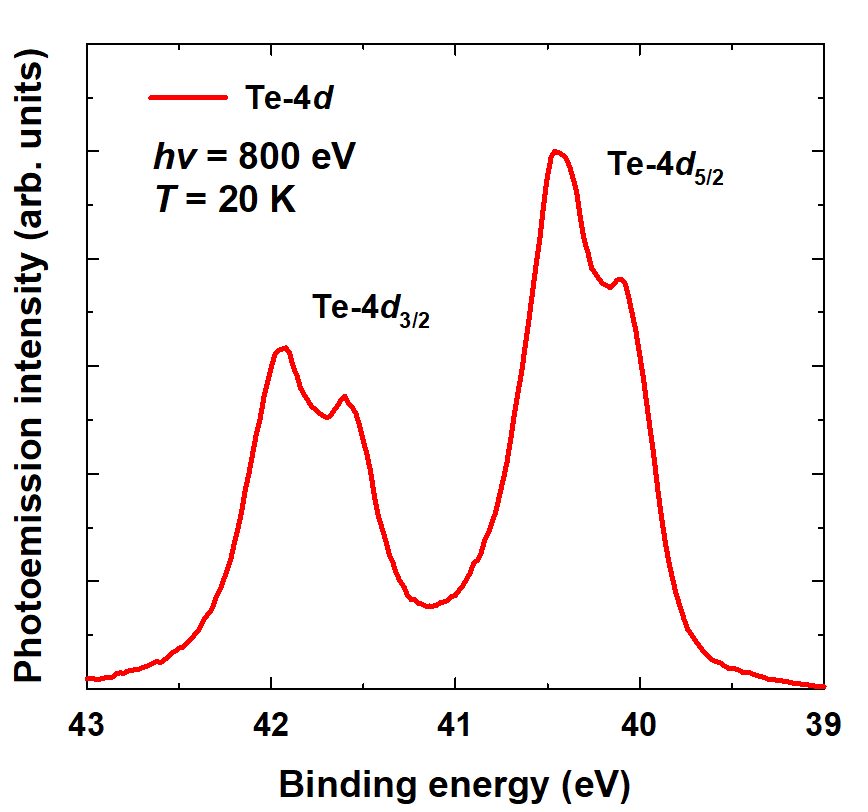}\,
\caption{Te-4\textit{d} core-level spectra of a single-crystalline UTe$_2$ measured at a photon energy of 800 eV, reproduced with permission from Shin-ichi Fujimori (seminar “Electronic structures of uranium compounds studied by photoelectron spectroscopy,” Charles University, May 2021). The spectra exhibit a clear splitting of the Te-4\textit{d} lines, in contrast to our Te-3\textit{d} spectra where no resolvable splitting is observed (Fig. 1, right). This comparison highlights differences in spectral features across core levels and measurement conditions.}
\label{fig:A4}
\end{figure}

%\nocite{*}
\clearpage
\bibliography{apssamp}

% Produces the bibliography via BibTeX.

\end{document}